\documentstyle[12pt]{article}
\def\ba{\begin{eqnarray}}
\def\ea{\end{eqnarray}}

\def\lb{\label}
\def\be{\begin{equation}}
\def\ee{\end{equation}}

\begin{document}
\title{Killing-Yano tensors and some applications}
\author{Osvaldo P. Santillan \thanks{Departamento de Matematica, FCEyN, Universidad de Buenos Aires, Buenos Aires, Argentina
firenzecita@hotmail.com and osantil@dm.uba.ar}}
\date {}
\maketitle
\begin{abstract}
The role of Killing and Killing-Yano tensors for studying the geodesic motion of the particle and the superparticle in a curved background is reviewed.
Additionally the Papadopoulos list \cite{Papadopoulos1} for Killing-Yano tensors in $G$ structures is reproduced by studying the torsion types these structures admit. The Papadopoulos list deals with groups $G$ appearing in the Berger classification, and we enlarge the list by considering additional $G$ structures which are not of the Berger type. Possible applications of these results in the study of supersymmetric particle actions and in the AdS/CFT correspondence are outlined.

\end{abstract}
\tableofcontents
\section{Introduction}
Killing and Killing-Yano tensors \cite{Yano}-\cite{Somers} and their conformal generalizations \cite{Conformal KY}-\cite{Curto} are a powerful tool for studying General Relativity problems. When a given space time admits such a tensor, a classical constant of motion for probe particles appears. This is a reminiscent of the isometries, and it is often said that Killing and Killing-Yano tensors are the generators of hidden symmetries of the background. Additionally, the separability of the Hamilton-Jacobi equation for a particle moving in the Kerr space time \cite{CarterKerr}-\cite{Woodhouse} is closely related to the presence of a conformal Killing tensor of rank two. Furthermore this tensor admits an square root which is Killing-Yano \cite{Floyd}, \cite{Penrose} and which plays an important role in the separability of the Dirac equation in the rotating background \cite{Carter}.

At the quantum level Killing-Yano tensors are generators for non anomalous symmetries, while for Killing tensors this may not be the case. It is well known that when a particle in a given background is quantized then, for every globally defined Killing vector the background admits there corresponds an operator which commutes with the hamiltonian. But this assertion is false for Killing tensors in general, as the commutator of the corresponding operator with the hamiltonian may not vanish \cite{Carter}. Nevertheless, when a Killing tensor admits a square root which is Killing-Yano, the anomaly vanish identically \cite{CarterMclenagan}.

The similarities between the usual isometries and hidden symmetries discussed above raise the question wether or not Killing-Yano tensors do form an algebra. This issue was investigated in \cite{Kastor} and \cite{Kastor2} where it was argued that the natural generalization of the Lie bracket for Killing vectors is the Schouten-Nijenhuis bracket for Killing tensors. The outcome is that Killing-Yano tensors do not form a Lie algebra in general, at least with this particular operation, but they do when some extra conditions are satisfied. An example is the requirement for the metric to be of constant curvature. For Killing tensors instead, an associated graded algebra was reported in \cite{grado}.

The presence of hidden symmetries for a given background may give information about the algebraic type of the curvature. In four dimensions, the presence of a conformal and non degenerate Killing-Yano tensor of rank two in a generic space time implies that the curvature is of type D in the Petrov classification \cite{WalkerPenrose}-\cite{WalkerPenrose3}. The local form of the metrics is known explicitly \cite{Kinnersley}. The generalization of the Petrov classification to higher dimensions was performed in \cite{CMPP} \footnote{An extensive review about aspects of higher dimensional General Relativity can be found in \cite{ReallDurkee2} and references therein.} and this classification allowed the authors of \cite{Krtous1}  to prove that any space which admit a closed non degenerate conformal Killing-Yano tensor is of type D. This was based in previous work done in \cite{Oota1}. Furthermore, when the Einstein equations are imposed, these metrics become the Kerr-Taub-Ads family \cite{chen} which generalize the old Myers-Perry solution \cite{Myers}. But the converse of this statement is an open question, though some suggestions in this direction has been made in \cite{Mason}.

Soon after the appearance of \cite{Krtous1} the geodesic motion and the Hamilton-Jacobi and Dirac equations in these spaces was studied in \cite{Geodesicmotion}-\cite{Oota1}. The outcome is that both equations are separable. Additionally, the role of conformal Killing-Yano tensors for studying geodesic motion in double spinning black rings was pointed out in \cite{ReallDurkee1}, and a method for constructing conserved charges in asymptotically flat spaces by use of Killing-Yano tensors was given in \cite{AF} and in anti de Sitter space times in \cite{Jesi}.

Killing-Yano tensors also appear in other contexts of mathematical physics. For example, in the theory of gravitational instantons they are known to generate Runge-Lenz type symmetries \cite{Gibbonsruback}-\cite{Valent2}. The separability of the Dirac equation in the Kerr-Taub-Nut metrics was studied in \cite{Taubnutkerr}, and formal properties of Dirac operators for spaces with hidden symmetries were pointed out in several works such as \cite{Diracalldimensions}-\cite{Visinescudirac}. Furthermore, Killing-Yano tensors are generators for exotic supersymmetries in the spinning particle motion in a curved background \cite{Superparticula1}-\cite{Superparticula4}. These are symmetries which mixes bosonic and ferminonic coordinates but whose squares does not give the Hamiltonian as usual supersymmetries does  \cite{susyinthesky}-\cite{susyinthesky2}. Further research related to the motion of particles of abelian and non abelian charges in the presence of external fiels has been performed in \cite{Tanimoto}-\cite{Nersesian1} and these techniques were further applied to derive
N = 4 supersymmetric mechanics in a monopole background in \cite{Nersesian2}. The relation between Killing-Yano and integrable systems was subsequently studied  in \cite{Integrables1}-\cite{Integrables2}, and applications related to string movement were found in \cite{Japoneses1}-\cite{Turcos}.

Although their importance was understood for long ago, till recent times few examples of spaces admitting Killing-Yano tensors were known. This situation changed in the last years. The problem of find Killing-Yano tensors on spherically symmetric space times was studied in \cite{colins} and on pp-wave backgrounds in \cite{ppwave}. The Killing tensors for the Melvin universe were characterized in \cite{melvin}. The local form of certain lorenzian metrics admitting Killing-Yano tensors of higher order was studied in \cite{ferrando}, and the presence of hidden symmetries in the Plebanski-Demianski family was studied in \cite{Plebanski}.

Recently, the problem of classify the $G$ structures do admit Killing-Yano tensors was investigated by Papadopoulos in \cite{Papadopoulos1}. It is interesting to note that all the examples Papadopoulos is finding are Einstein or Ricci-flat. Furthermore, these spaces can be uplifted to an AdS supergravity solution and since the constant of motions of rotating string configurations in these backgrounds are related to quantum numbers in a conformal dual quantum field theory, the study of hidden symmmetries in these backgrounds may be of theoretical interest.

The present work is organized as follows. In section 2 the role of Killing and Killing-Yano tensors as generators for hidden symmetries for the particle and the spinning particle in a given space-time is reviewed. It is also emphasized the fact that when a Killing tensor has a Killing-Yano "square root" the classical symmetries it generates are non anomalous. In section 3 the conformal generalizations of Killing and Killing-Yano tensors and their role in finding solutions of a Dirac equations in the curved background are briefly described. In section 4 an attempt to generalize both notions for the motion of the Polyakov string and spinning string is presented. We are unable to find such a generalization unless some extra information about the string movement is given, and some examples of this situation are given explicitly. In section 5 the main features of $G$ structures and their relation to special holonomy manifolds are briefly discussed, and all the cases of the Papadopoulos list are reproduced by means of the torsion formalism developed in the references \cite{Agrikola1}-\cite{Karigiannis}. In addition we analize the presence of Killing-Yano tensors in almost contact structures and in $SO(3)$ structures in $SO(5)$ and in structures $H_k\subset SO(n_k)$, with $H_1=SO(3)$, $H_2=SU(3)$, $H_4=Sp(3)$ and $H_8=F_4$. Section 6 contains the discussion of the results and their possible applications.

\section{Killing-Yano tensors as exotic supersymmetries}

In the present section the most important aspects of Killing and Killing-Yano tensors and their role in for finding conserved quantities for motion of a particle and spinning particle in a curved background are reviewed. It is also emphasized the role of Killing-Yano as generators of exotic supersymmetries. Clear introductory notes are given for instance in \cite{Van Holten lectures}, and this reference can be consulted for further details.

\subsection{Killing tensors and the bosonic particle}

A bosonic particle falling freely in a geodesically complete background $(M, g_{\mu\nu})$ is described by the following action
\be\lb{lagro}
S=\int_{\tau_0}^{\tau_1}\textit{L}d\tau=\int_{\tau_0}^{\tau_1}g_{\mu\nu}(x)\dot{x}^\mu\dot{x}^\nu d\tau,
\ee
$\dot{x}^\mu=dx^\mu/d\tau$ being the derivative with respect to the proper time $\tau$ of the particle coordinate $x^{\mu}$. The variation of the action (\ref{lagro}) with respect to arbitrary infinitesimal transformations $\delta x$ and $\delta \dot{x}$ is
\be\lb{lagro2}
\delta S=\int_{\tau_0}^{\tau_1}\bigg[\frac{\delta \textit{L}}{\delta x^\mu}-\frac{d}{d\tau}\bigg(\frac{\delta \textit{L}}{\delta \dot{x}^\mu}\bigg)\bigg]\delta x^\mu d\tau+\int_{\tau_0}^{\tau_1}\frac{d}{d\tau}\bigg(\frac{\delta \textit{L}}{\delta x^\mu} \delta x^\mu\bigg) d\tau
\ee
$$
=\int_{\tau_0}^{\tau_1}\bigg[-\delta x^\mu g_{\mu\nu}\frac{D \dot{x}^\nu}{D\tau}+\frac{d}{d\tau}\bigg(\delta x^\mu p_\mu \bigg)\bigg]d\tau.
$$
$p_\mu$ being the momentum of the particle
\be\lb{momo}
p_\mu=g_{\mu\nu}\dot{x}^\nu.
\ee
 When the endpoints are fixed, i.e, when $\delta x^\mu=0$ the total time derivative in (\ref{lagro2}) may be discarded. Then variation (\ref{lagro2}) is zero when the Euler-Lagrange equations
\be\lb{ekmok}
\frac{D \dot{x}^\nu}{D\tau}=\ddot{x}^\mu+\Gamma^\mu_{\nu\alpha}\dot{x}^\nu \dot{x}^\alpha=0,
\ee
are satisfied. Here $\Gamma^\mu_{\nu\alpha}$ denote the usual Christoffel symbols constructed in terms of the metric $g_{\mu\nu}$
\be\lb{crist}
\Gamma_{ij}^k=\frac{g^{ik}}{2}(g_{ik,j}+g_{jk,i}-g_{ij,k}).
\ee
The first two members of the equations of motion (\ref{ekmok}) are the definition of the derivative $\frac{D \dot{x}^\nu}{D\tau}$. The vanishing of this derivative means that the particle moves along a geodesic line in the curved background.

 In the situations for which the variations $\delta x^\mu=K^\mu$ do not have fixed endpoints the total derivative in (\ref{lagro2}) can not be discarded. In this case, by taking (\ref{ekmok}) into account it follows that the  total variation of (\ref{lagro2}) is
\be\lb{sum}
\delta \textit{L}=\frac{d}{d\tau}\bigg(K^\mu p_\mu \bigg).
\ee
When $\delta x^\mu=K^\mu$ is such that this variation is zero it will be called a symmetry of $\textit{L}$ and for any symmetry the quantity
\be\lb{shro}
E_K= K_\mu \dot{x}^\mu,
\ee
is a constant of motion for the particle.

The most celebrated example of symmetries for (\ref{lagro2}) are those of the forms $\delta x^\mu=K^\mu(x)$. The condition for (\ref{sum}) to vanish gives
$$
\frac{d }{d\tau}\bigg(K_\mu \dot{x}^\mu \bigg) =\dot{x}^\nu\nabla_\nu K_\mu \dot{x}^\mu+K_\mu \frac{D \dot{x}^\mu}{D\tau}=0.
$$
But the last term is zero by (\ref{ekmok}) and the first one gives
\be\lb{killing}
\nabla_{(\nu} K_{\mu)}=0,
\ee
where the parenthesis denote the usual symmetrization operation. Equation (\ref{killing}) shows that the vector field $K_\mu$ is Killing, that is, a local isometry of $g_{\mu\nu}$. Thus for a particle moving along a geodesic in a given background $(M, g_{\mu\nu})$ there is a constant of motion for every isometry the background admits.

 The isometries considered above are not the whole set of symmetries. The most general ones are of the form $\delta x^\mu=K(x,\dot{x})$, that is, transformation which are local with respect to the phase space coordinates ($x^\mu$, $\dot{x}^\mu$). The generality of this ansatz follows from the fact that a dependence on higher order time derivatives such as $\ddot{x}$ will reduce to combinations of ($x$, $\dot{x}$) by means of the equations of motion (\ref{ekmok}) and thus it is redundant. If a Taylor like expansion of the form \be\lb{expans}
\delta x^\mu=K^\mu+K^\mu_\alpha \dot{x}^\alpha+K^\mu_{\alpha\beta} \dot{x}^\alpha \dot{x}^\beta+...,
\ee
with velocity independent tensors $K^\mu_{\mu_1..\mu_n}(x)$ is proposed, then a calculation analogous to the one leading to (\ref{killing}) shows that if (\ref{expans}) will be a symmetry of the lagrangian (\ref{lagro}) when
\be\lb{killingt}
\nabla_{(\mu} K_{\mu_1..\mu_n)}=0,
\ee
a condition which generalize (\ref{killing}). These tensors are known as Killing tensors and the quantities
\be\lb{canti}
c_n=K_{\mu_1..\mu_n}\dot{x}^{\mu_1}..\dot{x}^{\mu_n},
\ee
are constants of motion for the particle moving in the background. An obvious Killing tensor is the metric itself, that is, $K_{\mu\nu}=g_{\mu\nu}$. The corresponding conserved charge
\be\lb{jsrt}
H=\frac{1}{2}g^{\mu\nu}p_\mu p_\nu,
\ee
is the Hamiltonian for the particle.

 A remarkable difference between Killing vectors and Killing tensors is that the first generate symmetries even for the quantum version of (\ref{lagro2}), while for tensors an anomaly may appear. The simplest quantum version of the particle motion is obtained by replacing the momentum $p_{\mu}$ with the operator $\nabla_{\mu}$ and, for the scalar fields, the classical Hamiltonian (\ref{jsrt}) is replaced with the operator
\be\lb{jam2}
\widehat{H}=\nabla_\mu( g^{\mu\nu}\nabla_\nu),
\ee
which coincides the laplacian acting on scalar functions. Furthermore, for any vector field $K^\mu$ there corresponds a quantum mechanical operator $\widehat{K}=K^\mu\nabla_\mu$ for which the commutator with the hamiltonian is
\be\lb{conmu}
[\widehat{K}, \widehat{H}]=\frac{2-n}{n}K_{\mu}^{;\mu\nu}\nabla_{\nu}+K_{\mu}^{;\nu}\widehat{H}.
\ee
It is well known that when $K^{\mu}$ is Killing the corresponding quantum mechanical operator commute with the laplacian. This means
 that Killing vectors generate true quantum symmetries. The situation is different for Killing tensors. As an example, consider operators of the form $\widehat{K}_{(2)}=\nabla_\mu (K^{\mu\nu}\nabla_\nu)$. Then a lengthy calculation performed in \cite{Carter} \footnote{See also \cite{Visinescusigma} where some terms were corrected.} shows that
$$
[\widehat{K}_{(2)}, \widehat{H}]=2\nabla^{(\sigma}K^{\mu\nu)}\nabla_{\sigma}\nabla_{\mu}\nabla_{\nu}+3\nabla_{\mu}\nabla^{(\sigma}K^{\mu\nu)}\nabla_{\nu}\nabla_{\sigma}
$$
\be\lb{conmu2}
\nabla_{\sigma}\bigg(g_{\mu\nu}(\nabla^{\sigma}\nabla^{(\mu}K^{\nu\lambda)}-\nabla^{\lambda}\nabla^{(\mu}K^{\sigma\nu)})
+\nabla_{\nu}\nabla^{(\sigma}K^{\mu\lambda)}\bigg)\nabla_{\lambda}
\ee
$$
-\frac{4}{3}\nabla_{\nu}(R_{\mu}^{[\nu} K^{\sigma]\mu})\nabla_{\sigma}.
$$
In the situation in which $K^{\mu\nu}$ Killing tensor all the terms above will vanish except the last one. This can be paraphrased by saying that the classical symmetry that a Killing tensor generates will be anomalous, unless the integrability condition \be\lb{piaza}R_i^{[j} K^{k]i}=0,\ee is satisfied. These condition holds for instance when the metric is Einstein $R_{ij}=\Lambda g_{ij}$, in particular this is true for Ricci-flat metrics. This is also true when the Killing-tensor is the square $K_{\mu\nu}=f_{\mu}^{\alpha}f_{\alpha \nu}$ of a Killing-Yano tensor $f_{\mu\nu}$. The last situation will be discussed in the next subsections.

\subsection{Supersymmetric extension of the bosonic particle}

A supersymmetric generalization of the particle action (\ref{lagro}) is the spinning particle constructed in \cite{Superparticula1}-\cite{Superparticula4}. This was introduced as a suitable semi-classical approximation to the dynamics of a massive spin-1/2 particle such as the electron. Its construction involves a fermionic extension $M_{\xi}$ of the manifold $M$, which requires the introduction of a new set of Grassmann variables $\xi^\mu$ with $\mu=1,..,D$ with $D$ being the dimension of the background in which the particle lives. For a particle moving in an euclidean space with its flat metric $g=\delta_{ab}dy^a\otimes dy^b$ a supersymmetric extension is
\begin{equation}
L=\delta_{ab}(\dot{y}^{a}\dot{y}^{b}+
      {i\over2}\xi^{a}\dot{\xi}^b),
    \label{Lagrangian2}
\end{equation}
an action which is invariant under the supersymmetry transformations
\be\lb{amg3}
\delta y^{a}=-i \epsilon \;\xi^{a}, \qquad \delta \xi^{a}= \dot{y}^{a} \epsilon,
\ee
with $\epsilon$ being an anti-commuting (Grassmann) number. More precisely, the transformation given above induce a variation on the lagrangian which is proportional to a total time derivative and therefore it does not affect the equations of motion. The Euler-Lagrange derived from (\ref{Lagrangian2}) are
\be\lb{sopo}
\frac{d \dot{y}^{a}}{d\tau}=0,
\qquad
\frac{d\xi^a}{d\tau}=0.
\end{equation}
The meaning of the last equations is transparent, the first one shows that the bosonic coordinates parameterize a line and that the fermionic variables $\xi^{\mu}$ are constant in time.

The lagrangian (\ref{Lagrangian2}) and the supersymmetry transformations (\ref{amg3}) are referred to cartesian coordinates $y^a$. For curvilinear coordinates $x^\mu$ (such as polar ones) one may write the metric in an n-bein basis $e^a=\partial_\mu y^a dx^\mu$ as $g=\delta_{ab}e^a\otimes e^b$. Then in a new coordinate system $\xi^a$ defined through the relation $\xi^\mu=e^\mu_a\xi^a$ the action may be rewritten as
\begin{equation}
L=g_{\mu\nu}\dot{x}_{\mu}\dot{x}_{\nu}+
      {i\over2}g_{\mu\nu}\xi^\mu\frac{D\xi^\nu}{D\tau},
    \label{Lagrangian3}
\end{equation}
and the supersymmetry transformation becomes
\be\lb{amg4}
\delta x^{\mu}=-i \;\epsilon \;\xi^{\mu}, \qquad \delta \xi^{\mu}=\epsilon \;\dot{x}^{\mu}.
\ee
In the last equation the fermionic time derivative
\begin{equation}
\frac{D\xi^\mu}{D\tau}=\dot{\xi}^\mu+\dot{x}^\nu \Gamma^\mu_{\nu\lambda}\xi^\lambda,
\end{equation}
has been introduced. With this definition it is straightforward to check that the lagrangian (\ref{Lagrangian3}) is invariant under (\ref{amg4}). Furthermore, the fact that the curvature of the metric is trivial plays no role in this checking and thus the extension is valid for any metric $g_{\mu\nu}$. Therefore (\ref{amg4}) is a supersymmetric
 extension of the bosonic particle lagrangian (\ref{lagro}) in any background. The change of the action (\ref{Lagrangian3}) with respect to a variation $\delta x^\mu$ and $\delta \xi^a$ is
\be\lb{nvrv}
\delta S=\int d\tau \bigg[-\delta x^\mu \bigg(g_{\mu\nu}\frac{D\dot{x}^\mu}{D\tau}+i \xi^\lambda \xi^\kappa R_{\lambda\kappa\mu\nu} \dot{x}^\nu \bigg)+i\Delta \xi^\mu g_{\mu\nu}\frac{D\xi^\nu}{D\tau}
\ee
$$
+\frac{d}{d\tau}\bigg(\delta x^\mu p_{\mu}-\frac{i}{2}\delta \xi^\mu g_{\mu\nu}\xi^\nu\bigg)\bigg],
$$
where the momentum
\be\lb{varo1}
p_\mu=g_{\mu\nu}\dot{x}^\nu-\frac{i}{2}\Gamma_{\mu\nu\lambda}\xi^\nu\xi^\lambda,
\ee
has been introduced, together with the variations
\be\lb{varo2}
\Delta \xi^\mu=\delta \xi^\mu+\delta x^\nu \Gamma_{\nu \lambda}^\mu\xi^\lambda,
\ee
and the curvature tensor
\be\lb{curva}
R_{\mu\mu\lambda}^\kappa=\partial_\mu\Gamma_{\nu\lambda}^\kappa-\partial_\nu\Gamma_{\mu\lambda}^\kappa
+\Gamma_{\lambda\mu}^\rho\Gamma_{\rho\nu}^\kappa-\Gamma_{\lambda\nu}^\rho\Gamma_{\rho\mu}^\kappa.
\ee
The equations of motion derived from (\ref{Lagrangian3}) generalize (\ref{sopo}) and can be casted in the following form
\be\lb{sopo3}
\frac{D\xi^\mu}{d\tau}=0,
\qquad
\frac{D \dot{x}^{\mu}}{d\tau}=-\frac{i}{2}\xi^\lambda \xi^\kappa R_{\lambda\kappa\mu}^{\nu} \dot{x}_{\nu}.
\end{equation}
 The last (\ref{sopo3}) in fact can be rewritten in terms of the "spin tensor" $S^{ab}=\xi^a\xi^b$ as
\begin{equation}\label{sopol2}
\frac{D \dot{x}^{\mu}}{d\tau}=-\frac{i}{2}S^{ab} R_{ab\mu}^{\nu} \dot{x}_{\nu},
\end{equation}
which is analogous to the electromagnetic force with the tensor $S^{ab}$ replacing usual the electric charge as coupling constant. Additionally the first (\ref{sopo}) imply that
\be\lb{sopo32}
\frac{D S^{ab}}{D\tau}=0,
\ee
i.e, the tensor $S^{ab}$ is covariantly constant.

\subsection{Symmetries of the phase superspace}

 Given the spinning particle action (\ref{Lagrangian3}) the next task is to characterize its symmetries. By analogy with (\ref{expans}) one may consider a general symmetry transformation of the super phase space ($x$, $\dot{x}$, $\xi$). Higher order derivatives such as $\dot{\xi}$ should absent due to the equation of motion (\ref{sopo3}), which are of first order in time derivatives of $\xi$. The generalization of (\ref{expans}) in this situation is an expansion of the form
\be\lb{simtrasfv}
\delta x^\mu=K^\mu(x,\dot{x},\xi)=K^{(1)\mu}(x,\xi)+\sum_{n=1}^\infty \frac{1}{n!} \dot{x}^{\nu_1}....\dot{x}^{\nu_n}K_{\nu_1....\nu_n}^{(n+1)\mu}(x, \xi),
\ee
\be\lb{simtrasfv2}
\delta \xi^\mu=S^\mu(x,\dot{x},\xi)=S^{(0)\mu}(x,\xi)+\sum_{n=1}^\infty \frac{1}{n!} \dot{x}^{\nu_1}....\dot{x}^{\nu_n}S_{\nu_1....\nu_n}^{(n)\mu}(x, \xi).
\ee
The variation (\ref{nvrv}) when no endpoints are fixed vanish if and only if
\be\lb{simfas}
\frac{d}{d\tau}\bigg(\delta x^\mu p_{\mu}-\frac{i}{2}\delta \xi^\mu g_{\mu\nu}\xi^\nu\bigg)=0.
\ee
Note that for deducing these result the equations of motion (\ref{sopo3}) has been took into account. Denoting the quantity in parenthesis
(\ref{simfas}) as $M$, it follows from (\ref{simtrasfv})-(\ref{simtrasfv2}) that it has an expansion of the form
$$
M=\sum_{n=0}^\infty \frac{1}{n!} \dot{x}^{\nu_1}....\dot{x}^{\nu_n}M_{\nu_1....\nu_n}^{(n+1)\mu}(x, \xi),
$$
such that
\be\lb{erverver}
K^{(n)}_{\mu_1....\mu_n\nu}=M^{(n)}_{\mu_1....\mu_n\nu},\qquad n\geq 1
\ee
\be\lb{rever}
S^{(n)}_{\mu_1....\mu_n\nu}=i\frac{\partial K^{(n)}_{\mu_1....\mu_n}}{\partial\xi^\nu},\qquad n\geq 0.
\ee
Additionally, for an arbitrary function $M(x,\dot{x}, \xi)$ of the super phase space, a simple chain rule together with the equations of motions (\ref{sopo3}) shows that
\be\lb{timder}
\frac{dM}{d\tau}= \dot{x}^\mu\bigg(\frac{\partial M}{\partial x^{\mu}}-\Gamma^\nu_{\mu\lambda} (\dot{x}^\lambda\frac{\partial M}{\partial \dot{x}^{\nu}}+\xi^\lambda \frac{\partial M}{\partial \xi^\nu})-\frac{i}{2}\xi^\lambda \xi^\kappa R_{\nu\mu\lambda\kappa}\frac{\partial M}{\partial \dot{x}^{\nu}} \bigg).
\ee
With the use of (\ref{erverver})-(\ref{timder}) the following recurrence relations are obtained for $n\geq 1$
\be
K^{(n)}_{( \mu_1 ... \mu_n ;\mu_{n+1})} +\frac{\partial K^{(n)}_{(\mu_1 ... \mu_n }}{\partial\xi^\lambda} \Gamma_{ \mu_{n+1})\kappa}^{\lambda}\,
       \xi^{\kappa}\, \, =\,
        \frac{i}{2}\, \xi^\lambda \xi^\kappa R_{\lambda\kappa\nu(\mu_{n+1}}\,
       K^{(n+1)\nu}_{\mu_1 ... \mu_n)} ,\qquad n\geq 1.
\label{spr}
\ee
For $n=0$ one may define the quantity $K^{(0)}$ by the relation
\be\lb{revera}
S^{(0)}_{\mu}=i\frac{\partial K^{(0)}}{\partial\xi^\nu},
\ee
and the equation for $S^{(0)}_{\mu}$ is equivalent to
\be
K^{(0)}_{\;,\;\mu} +\frac{\partial K^{(0)}}{\partial\xi^\lambda} \Gamma_{ \mu\kappa}^{\lambda}\,
       \xi^{\kappa}\, \, =\,
        \frac{i}{2}\, \xi^\lambda \xi^\kappa R_{\lambda\kappa\nu\mu}\,
       K^{(1)\nu} .
\label{spra}
\ee
Note that different for the bosonic case, the scalar $K^{(0)}$ is not an irrelevant constant, because it may depend non trivially on $(x,\xi)$ by (\ref{spra}).
The equations (\ref{rever})-(\ref{spra}) characterize the local form of the symmetries for the superparticle action. These equations were derived, to the best of our knowledge, in the references \cite{susyinthesky}-\cite{susyinthesky2}. The deduction given on that references rely in the hamiltonian formalism, in which the symmetries are interpreted in terms of quantities which commute with the hamiltonian. The outcome is exactly the equations derived above. The hamiltonian formalism is suitable for generalizing the notion of hidden symmetries when the particle is in presence of gauge fields. This fact was exploited particularly in the references \cite{Tanimoto}-\cite{Nersesian2}.

\subsection{Exotic supersymmetries}

Although the equations (\ref{rever})-(\ref{spra})  given above characterize the symmetries of the action (\ref{Lagrangian3}) it may be very hard to find
explicit solutions for a given background. In the following some simple cases will be considered namely, the supersymmetries
already introduced in (\ref{amg4}) and the exotic supersymmetries generated by the Killing-Yano tensors.

The simplest solution of the system (\ref{rever})-(\ref{spra}) are symmetries which do not depend on the fermionic variables. In this case it is immediate to check that the resulting symmetry generators are Killing tensors. The simplest one is the metric tensor $g_{\mu\nu}$ and the associated conserved quantity is
\be\lb{jsrt2}
H=\frac{1}{2}g^{\mu\nu}p_\mu p_\nu,
\ee
with $p_\mu$ given in (\ref{varo1}). The last expression is the hamiltonian of the particle. In the Hamilton formalism the time evolution of any dynamical quantity $F(x,p,\xi)$ is given in terms of the Poisson bracket with the hamiltonian
\be\lb{evol}
\frac{dF}{d\tau} =\{ F, H \}.
\ee
The fundamental Poisson brackets of the theory (\ref{Lagrangian3}) are given by
\be\lb{efe2}
\{x^{\mu}, p_{\nu}\}= \delta_{\mu}^{\nu}, \qquad \{\xi^{\mu}, \xi^{\nu}\}=- i g^{\mu\nu}.
\ee
From these brackets it is straightforward to find that
\be\lb{efe3}
\left\{ p_{\mu}, \xi^\nu \right\}  =  \frac{1}{2} g^{\kappa\nu}g_{\kappa\lambda,\mu} \xi^\lambda,\qquad \left\{ p_{\mu}, p_\nu \right\}  =  -\frac{i}{4} g^{\kappa\lambda}g_{\kappa\rho,\mu}g_{\lambda\sigma,\nu} \xi^\rho\xi^\sigma.
\ee
In these terms, the following Poisson bracket for the tensor $S^{ab}$ are found
\be\lb{spin}
\left\{ S^{ab}, S^{cd}\right\}=\delta^{ad}S^{bc} +\delta^{bc}S^{ad} -\delta^{ac}S^{bd} -\delta^{bd}S^{ac},
\ee
which justify the name "spin tensor". The space like components
$S^{ab}$ represent the magnetic momentum and the time like components are the electric momentum. As it is expected that for free particles like
electrons that the electric momentum in the rest frame vanish identically, the time like components should vanish identically. This condition may be imposed by requiring by implementing the subsidiary condition
\be\lb{erv2}
\dot{x}^{\mu} \xi_{\mu}=0,
\ee
after solving the equations of motion \cite{susyinthesky}-\cite{susyinthesky2}.

 Another example of symmetries described by the system (\ref{rever})-(\ref{spra}) should be the supersymmetry transformations (\ref{amg4}), and it will be instructive to check this explicitly. By comparison between (\ref{amg4}) and (\ref{simtrasfv})-(\ref{simtrasfv2}) it is found that the non zero supersymmetry generators are
 \be\lb{susy}
 K^{(1)}_{\mu}=-i\;g_{\mu\nu}\xi^\nu, \qquad S^{(1)}_{\mu\nu}=g_{\mu\nu},
 \ee
 and the relation (\ref{rever}) is satisfied for all of them. Moreover one has that
 $$
 K^{(1)}_{\mu; \;\alpha}=g_{\mu\nu,\;\alpha}\;\xi^\nu-g_{\lambda \nu}\;\Gamma_{\mu\alpha}^\lambda \xi^\nu=g_{\mu\lambda}\Gamma_{\nu\alpha}^\lambda \xi^\nu,
 $$
 where in the last the equality it has been used that $g_{\mu\nu;\;\alpha}=0$. With the use of the last formula it is immediate that (\ref{spr}) is satisfied. In addition the left hand side of (\ref{spra}) is zero and by using the first (\ref{susy}) the right hand side vanish by the first Bianchi identity $R_{\mu[\nu\alpha\beta]}=0$. Thus the supersymmetry transformations (\ref{susy}) are solutions of the equations (\ref{rever})-(\ref{spra}), which gives an interesting consistency check. The conserved quantity related to the supersymmetry (\ref{amg4}) is obtained from (\ref{simfas}), the resulting Noether charge
\be\lb{suchar}
Q=p_\mu \xi^\mu,
\ee
is known as the supercharge.

One may consider, in addition to the above examples, symmetries which mimics the supersymmetry property of mixing bosonic and fermionic coordinates.
A natural ansatz for these symmetries is
\be\lb{like}
\delta x^{\mu}= - i \epsilon\, f^{\mu}_{a}(x)\; \xi^a.
\ee
If the 1-forms $f^{\mu}_a$ are an n-bein $e^{\mu}_a$ basis for the metric, then the previous formula will represent a true supersymmetry (\ref{amg4}). Otherwise it will be a new type of symmetry, whose composition does not necessarily close to the Hamiltonian. For this reason these are known as exotic supersymmetries. By comparing (\ref{simtrasfv})-(\ref{simtrasfv2}) with (\ref{like}) and taking into account (\ref{rever}) the following generator are obtained
\be\lb{ruber}
K^{(1)}_{\mu}=-i \; g_{\mu\nu}\;f^{\nu}_{a}(x)\; e^a_\alpha \;\xi^\alpha,\qquad S^{(1)}_{\mu\alpha}=g_{\mu\nu}\;f^{\nu}_{a}(x)\; e^a_{\alpha}.
\ee
In these terms the equation (\ref{spr}) is equivalent to
\be\lb{iqvo}
D_{\mu} f^{a}_{\nu}\, +\, D_{\nu} f^{a}_{\mu}\, =\, 0.
\ee
On the other hand, these can not be the whole generators. If this were the case
then the left hand side of the equation (\ref{spra}) will be zero, but the right hand side will not unless $f^{a}_{\nu}=e^{a}_{\nu}$. Thus, a non zero $K^{(0)}$ generator is present, and should be of the form
\be\lb{newge}
K^{(0)}=i \;c_{abc}\;\xi^a\xi^b\xi^c,
\ee
as the cubic dependence in $\xi^a$ follows by noticing that the right hand of (\ref{spra}) is multiplied by a quadratic expression in the $\xi^a$ variables and the generator $K^{(1)}_{\mu}$ in (\ref{ruber}) is linear in the Grassmann variables. With this new generator the equation (\ref{spra}) turns to be equivalent to
\be\lb{equo}
D_{\mu} c_{abc}\, =\, -\, R_{\mu\nu ab} f^{\nu}_{c}\, -\, R_{\mu\nu bc}
                  f^{\nu}_{a}\, -\, R_{\mu\nu ca} f^{\nu}_{b}.
\ee
In these terms the new symmetry transformations are
\be\lb{simtrasfvss}
\delta_f x^\mu= - i\; \epsilon\, f^{\mu}_{a}(x)\; \xi^a,
\ee
\be\lb{simtrasfv2ss}
\delta_f \xi^\mu=\epsilon\, f^{\mu}_{a}(x)\; e^a_{\nu}\;\dot{x}^\nu+\epsilon \; c_{\mu\nu\alpha}\;\xi^\nu\xi^\alpha.
\ee
A further simplification is obtained with the requirement that the transformations $\delta_f$ anti-commute with the supersymmetry transformation $\delta_s$
\be\lb{suin}
\{\delta_f, \delta_s \} = 0.
\ee
Such transformations are called superinvariant and the equation (\ref{suin})
imply that
\be\lb{noanti}
f_{\mu}^{a} e_{\nu a}+f_{\nu}^{a} e_{\mu a}=\, 0.
\ee
The last equation means that the tensor $f_{\mu\nu}\, =\, f_{\mu}^{a} e_{\nu a}$ is completely antisymmetric. The equations (\ref{iqvo}) are in this case equivalent to the following one
\be\lb{kiya}
f_{\mu\nu ;\lambda} + f_{\lambda\nu ;\mu}\, =\, 0.
\ee
Tensors satisfying (\ref{kiya}) are known as Killing-Yano tensors. In brief, Killing-Yano tensors generate superinvariant exotic supersymmetries.
By taking into account (\ref{kiya}) and the complete antisymmetry of $f_{\mu\nu}$ it follows that the gradient
$$
 f_{\mu\nu;\lambda}=\frac{1}{3}( f_{\mu\nu ;\lambda} + f_{\nu\lambda ;\mu}
                            + f_{\lambda\mu ;\nu} )
=H_{\mu\nu\lambda} ,
$$
is completely antisymmetric and thus it defines a three form
$H_{\mu\nu\lambda}$. By taking the second covariant derivative of the last equation
together with the Ricci identity and the antisymmetry of $f_{\mu\nu}$ gives that
\be\lb{rivo2}
H_{\mu\nu\lambda ;\kappa}\, =\, \frac{1}{2}\, \bigg( R_{\mu\nu\kappa}^{\sigma}
   f_{\sigma\lambda} + R_{\nu\lambda\kappa}^{\:\:\:\:\:\:\:\sigma} f_{\sigma\mu} +
   R_{\lambda\mu\kappa}^{\:\:\:\:\:\:\:\sigma} f_{\sigma\nu} \bigg).
\ee
By comparison of (\ref{rivo2}) with (\ref{equo}) one get the following identification
\be\lb{rivo3}
c_{abc}= - 2 H_{abc} =- 2 e^{\mu}_a e^{\nu}_b e^{\lambda}_c H_{\mu\nu\lambda}.
\ee
In conclusion, the most general supersymmetry like symmetries of the form (\ref{like})
are obtained by the generators of the form (\ref{simtrasfvss})-(\ref{simtrasfv2ss}) and if these symmetries are superinvariants in the sense of (\ref{suin}) then these symmetries are completely determined in terms of Killing-Yano tensors of second rank, which are antisymmetric tensors satisfying (\ref{kiya}). In these terms the exotic supersymmetry is defined by the formulas (\ref{suin}),(\ref{noanti}) and (\ref{rivo3}) \cite{susyinthesky}-\cite{susyinthesky2}.

\subsection{Squares of exotic symmetries}

In the hamiltonian formalism, where the fundamental brackets are (\ref{efe2})-(\ref{efe3}), the action of symmetry transformation over a function of the super phase space $F(x,p,\xi)$ is given as
\be\lb{secv}
\delta F =i \{ F, Q_s\}\epsilon,
\ee
$Q_s$ being the conserved constant of motion. In particular, it can be checked that (\ref{simtrasfv})-(\ref{simtrasfv2}) is direct consequence of (\ref{secv}) together with the definition of the supercharge (\ref{suchar}) and the fundamental Poisson bracket (\ref{efe2})-(\ref{efe3}), which gives a consistency check. By using (\ref{efe3}) it is seen that the supersymmetry generator $Q$ satisfy
\be\lb{wk}
\{Q, Q\}=-2\;i\;H,
\ee
which is a well known feature of the supersymmetry transformations.

 The discussion of the previous paragraph may be generalized as follows. When $r$ symmetries transformations $\delta_i$ with $i=1,..,r$ are present, then there exist $r$ conserved supercharges $Q_i$  defined by (\ref{simfas}). Let us denote as $Z_{ij}$ the following Poisson bracket
\be\lb{tobe}
\{ Q_i, Q_j \} =- 2\;i \;Z_{ij}.
\ee
The time derivative of this quantity is
\be\lb{erefere}
\frac{dZ_{ij}}{d\tau}=\{H, Z_{ij} \}=- 2\;i \;\{H, \{ Q_i, Q_j \} \}=
=- 2\;i \;\{Q_j, \{ H, Q_i \} \}- 2\;i \;\{Q_i, \{ Q_j, H \} \}=0,
\ee
where in the last step the Jacobi identity together with the fact that $\{ Q_i, H \}=0$ had been used. The quantity $Z_{ij}$ is the "charge" corresponding to the transformation $\delta_{ij}=\{\delta_i,\delta_j\}$ and the relation (\ref{erefere}) imply that $\delta_{ij}$ is a symmetry transformation as well. In particular, if the symmetries $\delta_i$ are exotic supersymmetries of the form (\ref{simtrasfvss})-(\ref{simtrasfv2ss}) then
\be\lb{erfer}
\delta_{ij}\, x^\alpha=\,  K^{\alpha\mu}_{ij}\;\dot{x}_{\mu}  +
             \frac{i}{2}\;I^{\alpha}_{ij ab}\;\xi^a\xi^b,
\ee
\be\lb{froto}
\delta_{ij}\, \xi^a = i\; I^{a \mu}_{ij b}\;\dot{x}_{\mu} \xi^b  -  G_{ij bcd}^a\;\xi^b\xi^c\xi^d,
\ee
the new quantities being defined as
$$
K^{\mu\nu}_{ij}  =  K^{\nu\mu}_{ij}\, =\, \frac{1}{2}\,\bigg( f^{\mu}_{ia}
                      f^{\nu a}_j + f^{\mu}_{j\:a} f^{\nu a}_i \bigg),
$$
\be\lb{dufono}
I^{\mu}_{ij ab}  =  \bigg( f^{\nu}_{i\:b} D_{\nu}
                   f^{\mu}_{j\:a}\, +\, f^{\nu}_{j\:b} D_{\nu} f^{\mu}_{i\:a}\,
                   +\, \frac{1}{2} f^{\mu c}_i c_{j\:abc}\, +\, \frac{1}{2}
                   f^{\mu c}_j c_{i\:abc}  \bigg),
\ee
$$
G_{ijabcd} =  \bigg( R_{\mu\nu ab}
             f^{\mu}_{i c} f^{\nu}_{jd}\, +\,
             \frac{1}{2} c_{i\:ab}^{e} c_{jcde} \bigg).
$$
The Killing Yano equations (\ref{rever})-(\ref{spra}) for $f^{\nu}_{jd}$ and $c_{abc}$ imply the following relations for the new quantities \cite{susyinthesky}-\cite{susyinthesky2}
$$
K_{ ( \mu \nu; \lambda )} = 0,
$$
\be\lb{mutrox}
D_{( \mu} I_{ \nu) ab} =R_{ab( \mu}
K_{\nu)},
\ee
$$
D_{\mu} G_{abcd} = R_{\lambda\mu [ab}I^{\lambda}_{cd]}.
$$
Note that the first (\ref{mutrox}) shows that the entries of the matrix $K_{ ij \mu \nu; \lambda }$ are all Killing tensors. This result is well know, the "square" of two Killing-Yano tensors gives a Killing tensor, a result which was obtained in General Relativity in \cite{Penrose}. Furthermore, it can be shown by taking into account (\ref{kiya}) that this Killing tensor satisfies
the integrability condition (\ref{piaza}) and therefore give rise to a symmetry which is free of anomalies, a result that was anticipated by Carter in \cite{Carter}, \cite{CarterMclenagan}.

\section{Conformal generalizations of Killing and Killing-Yano tensors}

The relations described above between Killing and Killing-Yano tensors can be generalized to tensors of higher order,
and to conformal generalizations \cite{nikitin}. The conformal generalization of a Killing vector is a vector field $K$ which
satisfy
$$
L_K g_{\mu\nu}=\lambda g_{\mu\nu},
$$
with $\lambda$ being a constant. This are vectors with a flow preserving a given conformal class of metrics. Clearly, when $\lambda$ is zero we recover the usual definition of a Killing vector. Similarly a conformal Killing tensor is
\be\lb{ckt}
\nabla_{(\nu}K_{\mu_1...\mu_n)}=g_{\nu(\mu_1}\widetilde{K}_{\mu_2...\mu_n)},
\ee
with $\widetilde{K}_{\mu_2..\mu_n}$ is the tensor defined by taking the trace on both sides.

Killing-Yano tensors also admit a generalization to orders higher than two, and conformal generalizations \cite{Conformal KY}-\cite{Curto}. To see this
one may note that the equation (\ref{kiya}) defining Killing-Yano tensors may be rewritten as
\be\lb{cherri}
\nabla_X f=\frac{1}{p+1}i_X df,
\ee
with $p=2$ and $X$ an arbitrary vector field. For an arbitrary p-form we will say that is Killing-Yano if (\ref{cherri}) is satisfied. The conformal generalization are known as conformal Killing-Yano tensors $f$ and are defined by the following equation \cite{Conformal KY}-\cite{Curto}
\be\lb{cucho}
\nabla_X f=\frac{1}{p+1}i_X df-\frac{1}{n-p+1}X^{\flat}\wedge d^{\ast}f.
\ee
Here $X^{\flat}$ is the dual 1-form to the vector field $X$ and the operation $d^{\ast} f=(-1)^p\ast^{-1} d \ast f$ has been introduced, in which
$$
\ast^{-1}=\epsilon_p \ast,\qquad \epsilon_p=(-1)^p\frac{\det g}{|\det g|}.
$$
Note that if $d^{\ast}f=0$ the CKY tensor reduces to a usual KY tensor. In terms of two CKY tensors of the same order one may construct a symmetric
two tensor
\be\lb{peqqu}
K_{\mu\nu}=(f^1)_{\mu \mu_1..\mu_n}(f^2)_{\nu}^{\mu_1..\mu_n},
\ee
which by virtue of (\ref{cherri}) is a conformal Killing tensor. In particular if the $f^i$ are Killing-Yano then (\ref{peqqu}) will be
Killing, and this generalize the fact indicated in (\ref{mutrox}) and found in \cite{Carter}. A particular case exploited in the literature are principal conformal Killing-Yano tensors for $p=2$. This are closed, i.e, $df=0$ and non degenerate and satisfy the equation
\be\lb{principo}
\nabla_X f=X^{\flat}\wedge \xi^{\flat},\qquad \xi_{\nu}=\frac{1}{n-1}\nabla_{\mu} f^{\mu}_{\nu}.
\ee
The vector $\xi_{\mu}$ satisfy the following equation
$$
\xi_{(\mu;\nu)}=-\frac{1}{n-2}R_{\lambda(\mu}f_{\nu)}^{\lambda},
$$
and it follows that for Ricci-flat or Einstein space this vector will be Killing. These tensors were considered in \cite{Krtous1} and it was proved in that reference that any space admitting a conformal and principal Killing-Yano tensor of order two is of type D in the classification of \cite{CMPP}.
Furthermore, when the Einstein equations are imposed, these metrics become the Kerr-Taub-Ads family \cite{chen}. Higher dimensional Killing-Yano tensors were considered in the context of black holes physics in \cite{higher}.

\subsection{Quantum symmetries from Killing-Yano tensors}
In view of the results discussed in previous sections, Killing-Yano tensors
are more fundamental than Killing tensors as they generate true symmetries for the movement of the free particle in a given curved background. In other words, they generate operators which commute with the wave operator on the curved background. An additional property, which makes them specially interesting, is
that they also generate operators which commute with the Dirac operator for the given background, thus they generate quantum symmetries for spin $1/2$ particles moving in the space time \cite{Diracalldimensions}-\cite{Visinescudirac} (see also \cite{Visinescusigma}). Recall that the standard Dirac operator on a curved background is defined as
\be\lb{dirc}
D=e^a \nabla_{X_a},
\ee
with $e^a$ a tetrad basis for the metric $g_{ab}$ of the background
$$
e^a e^b+e^b e^a=g^{ab}.
$$
In these terms one may construct the following operators acting on spinors
\be\lb{volv}
D_f=L_f+(-1)^p f D,
\ee
with
$$
L_f=e^af\nabla_{X^a}+\frac{p}{p+1}df+\frac{n-p}{n-p+1}d^\ast f,
$$
being an operator constructed in terms of a p-form whose components are $f_{\mu_1...\mu_p}$. The graded commutator
$$
\{D, D_f\}=D D_f+(-1)^p D_f D,
$$
calculated between the operators (\ref{volv}) and (\ref{dirc}) is given by
\be\lb{sevvv}
\{D, D_f\}=R D, \qquad
R=\frac{2(-1)^p}{n-p+1}d^{\ast} f D.
\ee
For a Killing-Yano tensor one has that $d^{\ast}f=0$ and thus $R=0$. This means that there exist operator for which the graded commutator with
the Dirac operator is zero for every Killing-Yano tensor the background admits. These properties were extensively studied for instance in \cite{Cariglia}.

\section{Killing-Yano tensors in string and superstring backgrounds}

\subsection{Hidden symmetries for the bosonic string}
As Killing and Killing-Yano tensors are generators for hidden symmetries for the particle and superparticle one may ask if there exist the analogous structures
for the movement of a Polyakov string in a given background. As is well known, the Polyakov action for the string is constructed as follows. Consider a $D$-dimensional space time $M$ with metric $g_{\mu\nu}$
a two dimensional worldsheet $\Sigma$ parameterized by coordinates $(\sigma^1, \sigma^2)$, and suppose that there is an embedding $\phi$ from $\phi: \Sigma\to M$ such that $x^\mu=x^\mu(\sigma_i)$. The Polyakov action is then expressed as
\be\lb{Pol}
S_{p}=T \int d^2\Sigma \sqrt{h}h^{ab}g_{\mu\nu}\partial_a x^\mu \partial_b x^{\nu},
\ee
where $h_{ab}$ is a metric in the two dimensional worldsheet $\Sigma$. It can be seen that by use of the equation of motion of $h^{ab}$ and replacing the result into (\ref{Pol}) one obtains the Nambu-Goto string. If we denote $\sigma^1=\tau$, $\sigma^2=\sigma$, $\dot{x}^\mu=\partial_\tau x^\mu$ and $x'^{\mu}=\partial_\sigma x^\mu$ then the Nambu-Goto action reads
\be\lb{NG}
S_{NG}=-T \int d^2\Sigma \sqrt{(g_{\mu\nu}\dot{x}^\mu x'^{\nu})^2-(g_{\mu\nu}\dot{x}^\mu \dot{x}^\nu)(g_{\mu\nu}x'^\mu x'^\nu)}.
\ee
The Polyakov action (\ref{Pol}) is invariant under diffeomorphisms and Weyl transformations $h^{ab}\to \Omega^2 h^{ab}$. The fields of the theory
are the bosonic coordinates $x^\mu$ and the three components $h^{ab}$ of the Riemann surface metric, which depends functionally on the two coordinates
$(\sigma^1, \sigma^2)$ parameterizing the surface.

 In order to find the general symmetries for the Polyakov action one should consider a variation $\delta \sigma$ and $\delta \tau$ and field variations $\delta h^{ab}$  and $\delta x^\mu$ of the bosonic fields defined at the same point. The total variation of the bosonic fields is
\be\lb{totvar}
\Delta h^{ab}=\delta h^{ab}+ \partial_i h^{ab}\delta \sigma^i,\qquad \Delta x^{\mu}=\delta x^{\mu}+ \partial_i x^\mu \delta \sigma^i.
\ee
The variation of the action (\ref{Pol}) with respect to (\ref{totvar}) inside a region $R$ is given by
\be\lb{varak}
\delta S=\int_R d^2\Sigma \;\bigg\{\bigg[\frac{\delta \textit{L}}{\delta x^\mu}-\partial_a\bigg(\frac{\delta \textit{L}}{\delta(
\partial_a x^\mu)}\bigg)\bigg]\delta x^\mu + \frac{\delta \textit{L}}{\delta h^{ab}}\delta h^{ab}\bigg\}
\ee
$$
+\int_{R}d^2\Sigma \;\partial_a\bigg(\frac{\delta \textit{L}}{\delta(\partial_a x^\mu)}\delta x^\mu +\textit{L}\delta\sigma^a\bigg).
$$
Explicitly this variation is
 $$
\delta S=\int_R d^2\Sigma \;\bigg[2\sqrt{h}h^{ab}g_{\nu\kappa,\mu}\partial_a x^\nu \partial_b x^{\kappa}-\partial_a\bigg(\sqrt{h}h^{ab}g_{\mu\kappa}\partial_b x^{\nu}
\bigg)\bigg]\delta x^\mu
$$
\be\lb{varak2}
+ \int_R d^2\Sigma\;\sqrt{h}(g_{\nu\kappa}\partial_a x^\nu \partial_b x^{\kappa}-\frac{1}{2}h_{ab}h^{cd}g_{\nu\kappa}\partial_c x^\nu \partial_d x^{\kappa})\;\delta h^{ab}
\ee
$$
+\int_{R}d^2\Sigma \;\partial_a\bigg(\sqrt{h}h^{ab}g_{\nu\kappa}\partial_b x^{\kappa}\Delta x^\nu \bigg).
$$
The Euler lagrange equations are obtained by considering variations that vanish on the boundary $\partial R$ of the region $R$. For Riemann surfaces in $0$ genus one may brought
$h_{ab}$ to a diagonal metric $\eta_{ab}$ by a conformal transformation. The equations of motion then are
\be\lb{eqkver}
 \eta^{ab}\partial_a \partial_b x^{\nu}+\eta^{ab}\Gamma^\kappa_{\mu\nu} \partial_a x^\mu\partial_b x^{\nu} =0,
\ee
which generalize the geodesic equation for a two dimensional motion. Alternatively, the last system of equations may be expressed as
\be\lb{eqkver2}
 \eta^{11}\frac{D \dot{x}^\mu}{D\tau}+
\eta^{22}\frac{D x'^{\mu}}{D\sigma}=0,
\ee
and the conformal constraints reduce to
$$
g_{\nu\kappa}(\dot{x}^\nu  x'^{\kappa}+ x'^\nu \dot{x}^{\kappa})=0,                                                                                                                      $$
\be\lb{confor}
\eta^{11}
g_{\nu\kappa}\dot{x}^\nu \dot{x}^{\kappa}+\eta^{22}g_{\nu\kappa} x'^\nu  x'^{\kappa}=0.
\ee
If instead one consider coordinate dependent variations $\delta x^{\mu}=K^{\mu}$ which do not vanish on the boundary and which leave the action invariant, then the vanishing of the variation (\ref{varak2}) together with the equations of motion (\ref{eqkver2}) imply that
$$
\eta^{11}\partial_\tau(\dot{x}^{\mu} K_\mu)+\eta^{22}\partial_\sigma(x'^{\mu} K_\mu)=0.
$$
By use of the equation of motions (\ref{eqkver2}) the last formula reduce to
\be\lb{fono}
\eta^{11}\dot{x}^\nu \dot{x}^{\mu}\nabla_{(\nu} K_{\mu)}+\eta^{22}x'^\nu x'^{\mu} \nabla_{(\nu} K_{\mu)}=0.
\ee
By comparing this with the second (\ref{confor}) it follows directly the following solution of this equation
\be\lb{ckv}
\nabla_{(\nu} K_{\mu)}=\lambda g_{\mu\nu},
\ee
$\lambda$ being an arbitrary constant. For $\lambda=0$ the vector $K_\mu$ is Killing, otherwise it is a conformal Killing vector.
Thus conformal Killing vectors generate constants of motion for the Nambu-Goto string.

In order to search for generalizations of Killing tensors for the Polyakov string one may postulate a symmetry transformation which depends also on the world sheet derivatives of the background
coordinates, that is, $\delta x^\mu=K^{\mu}(x, \cdot{x}, x')$. Then by performing a Taylor like expansion of the form
\be\lb{qt}
\delta x^\mu=K^\mu+K^\mu_{\nu\alpha} \dot{x}^\nu x'^\alpha +...
\ee
the vanishing of the action (\ref{varak2}) gives the following system to solve
$$
\eta^{11}\partial_\tau(\dot{x}^{\mu} \dot{x}^\nu x'^\alpha K_{\mu\nu\alpha}  )+\eta^{22}\partial_\sigma(x'^{\mu}  \dot{x}^\nu x'^\alpha K_{\mu\nu\alpha})=0.
$$
Unfortunately, we find very difficult to solve this system. Recall that the main task is to find a geometrical object which give rise to a conserved quantity
 for any solution of the equation of motions. But we have found that, due to the mixing of derivatives in $\tau$ and $\sigma$, in order to have a conserved
quantity one should impose additional conditions on the equations of motion. This failure suggest that in order to find a hidden symmetry for an string movement, one should partially specify the way that the string evolves. For instance, one may be studying an spinning or a rotating string, or other similar configurations like a wound string, and after specifying this behavior one may search for hidden symmetries. A simple example of this is to consider the Nambu-Goto string (\ref{NG}) which can be rewritten as
\be\lb{NG2}
S=\int_{\sigma}d^2\Sigma\sqrt{-\det(g_{\mu\nu}\partial_a x^\mu\partial_b x^\nu)}.
\ee
If the background metric $g_{\mu\nu}$ admits a globally defined Killing vector field $V$ then one may rewrite the induced metric $\hat{g}_{\mu\nu}$ on $M/G$, with $G$ the orbits of the Killing vector,
as follows
\be\lb{asv}
\hat{g}_{\mu\nu}=g_{\mu\nu}-\frac{\xi^\mu\xi^\nu}{g_{00}}.
\ee
If additionally the string world surface is foliated by the orbits $G$ of the Killing vector \cite{Japoneses1} then the Nambu-Goto action reduce to
\be\lb{be}
S=\int_{\sigma_0}^{\sigma_1}\bigg(\frac{1}{N}\widetilde{g}_{\mu\nu}(x)x'^\mu x'^\nu+N\bigg)\; d\sigma
\ee
where we have defined a lapse function $N$ which under a reparameterization $\sigma'=\sigma'(\sigma)$ transforms as
\be\lb{lapse}
N\to N'=\frac{d\sigma}{d\sigma'}N
\ee
in order to insure the action (\ref{be}) to be invariant under reparameterizations. Here we have denoted $\widetilde{g}_{\mu\nu}=g_{00}\hat{g}_{\mu\nu}$. From here it follows
that when the string world surface is foliated by the orbits of a Killing vector then the action reduce to a one dimensional action with an effective metric $\widetilde{g}_{\mu\nu}=g_{00}\hat{g}_{\mu\nu}$ \cite{Japoneses1}. Clearly the Killing and Killing-Yano this metric admits will generate hidden symmetries for the motion, as in the particle case.

\subsection{Hidden symmetries for the spinning string}

Considerations analogous to the above hold for the movement of the spinning string, whose action in the conformal gauge is
\begin{equation}
S=\int d\sigma^2 \bigg(\eta^{ab}g_{\mu\nu}\partial_a x^{\mu}\partial_b x^{\nu}+
      {i\over2}g_{\mu\nu}\xi^\mu\rho^a \frac{D\xi^\nu}{D\sigma_a}\bigg),
    \label{Lagrangianc3}
\end{equation}
and is supplemented with the vanishing on the worldsheet of the energy momentum tensor $T_{ab}$
\be\lb{emom}
T_{ab}=g_{\mu\nu}\partial_a x^{\mu}\partial_b x^{\nu}+
      {i\over2}g_{\mu\nu}\psi^\mu\rho^{(a} \frac{D\psi^\nu}{D\sigma^{b)}}-\frac{\eta^{ab}}{2}\bigg(g_{\mu\nu}\partial_c x^{\mu}\partial^c x^{\nu}+
      {i\over2}g_{\mu\nu}\psi^\mu\rho^{c} \frac{D\psi^\nu}{D\sigma^{c}}\bigg)=0
\ee
and the supercharge $Q_a$
\be\lb{scr}
Q_a=\frac{1}{2}\rho^b \rho^a \psi_\mu D_a x^\mu=0.
\ee
In presence of a Killing vector $V^\mu$ the induced metric on $M/G$ is(\ref{asv}). A way to reduce the action to a particle one is to assume that the spinning string movement is foliated by the orbits of the Killing vector. Also the further requirement
\be\lb{fr}
V_\mu \partial_\sigma x^\mu=0,\qquad \pounds_V \psi^\mu=0,
\ee
$$
\psi^\mu V_\mu=g_{00}\Upsilon,
$$
$\Upsilon$ being a constant spinor,  implies the decomposition $\psi^\mu=\xi^\mu+V^\mu \Upsilon$. Under these assumptions the action (\ref{Lagrangianc3}) reduce to $S=I \Delta \tau$ with
$$
I=\int d\sigma \bigg(\widetilde{g}_{\mu\nu}\dot{x}_{\mu}\dot{x}_{\nu}+
      {i\over2}\widetilde{g}_{\mu\nu}\xi^\mu\frac{D\xi^\nu}{D\sigma}\bigg),
$$
with the dots denoting derivatives with respect to $\sigma$ \cite{Turcos}. The last expression is equivalent to (\ref{Lagrangian3}) with the induced metric $\widetilde{g}_{\mu\nu}$. We have then reproduced the particle limit of the spinning string found in \cite{Turcos} and the Killing-Yano tensors the induced metric admits will generate hidden symmetries for the motion of this configuration of the spinning string.

\section{Killing-Yano tensors and G-structures}

 Our next task is to investigate the presence of Killing-Yano tensors in $G$ structures. These structures play an important role for constructing supergravity solutions and appear naturally when studying special holonomy manifolds. As is well known, the holonomy group of a metric $g$ defined over an oriented $n$-dimensional manifold $M$ may be $SO(n)$ or a subgroup $G\in SO(n)$. The possible holonomy subgroups were classified by Berger in \cite{Berger}. The groups we will be concerned with are $Spin(7)$, $G_2$, $Sp(n)$, $Sp(n)\times Sp(1)$, $U(n)$ and $SU(n)$ and it turns out that metrics with these holonomy groups are always Einstein or Ricci flat. For the Ricci flat case, the reduction of the holonomy to $G$ is equivalent to the presence of a set of p-forms, which will be denoted from now as $\sigma^G_p$, which are constructed in terms an n-bein basis $e^a$ for $g$ and each of which is invariant under the action of $G$ and also covariantly constant with respect to the Levi-Civita connection. The situation is a bit different for the Einstein case, as we will see below.

To give an example, consider a 7-metric $g_7=\delta_{an}e^b\otimes e^b$ with $e^a$ a 7-bein basis and $a,b=1,..,7$. Then the following three form
\be\lb{seven}
\phi=c_{abc}e^a\wedge e^b\wedge e^c,
\ee
constructed in terms of the multiplication constants $c_{abc}$ of the imaginary octonions, is invariant under a $G_2$ rotation of the basis $e^a$. This follows from the fact that $G_2\in SO(7)$ is the automorphism group of the imaginary octonions. The set composed by the metric $g_7$ and the 3-form (\ref{seven}) is called a $G_2$ structure. In general s $G$ structure is composed by a riemannian metric $g$ together with a complete set of $G$ invariant p-forms $\sigma_p^G$. For $G=G_2$ the additional condition $\nabla_X \phi=0$ for an arbitrary vector field $X$ implies that the parallel transport of the $e^a$ around a closed loop will induce a rotation $e'^a=R^a_b e^b$ which leaves $\phi$ invariant. Thus in this case the holonomy will be $G_2$ or a subgroup of $G_2$. The resulting equations are equivalent to the differential system $d\phi=d\ast \phi=0$ \cite{Bryant}. Similar consideration follows for Ricci flat $G$ structures. The condition $\nabla \sigma^G_p=0$ will imply that the holonomy is reduced to $G$ or to a smaller subgroup.

An important tool for studying $G$ holonomy manifolds is the torsion formalism, which is a method for studying obstructions for a metric $g$ to be of $G$ holonomy and was reviewed in \cite{Agrikola1}. The roots of this formalism dates, to the best of our knowledge, from the work \cite{Obata} about hypercomplex structures, and it can briefly be described as follows. The Berger holonomy groups $G$ are embedded in $SO(n)$ and this imply algebra $so(n)$ can be represented schematically as $so(n)=g\oplus g^{\perp}$. For Ricci-flat holonomy groups, this induce the following decomposition of the Levi-Civita connection
\be\lb{schemo}
\nabla=\nabla^g+\nabla^{g^\perp}=\nabla^g+\frac{1}{2}T,
\ee
the component $\nabla^g$ satisfying $\nabla_X\sigma_p^g=0$. The equality (\ref{schemo}) can be taken as the definition of the torsion tensor $T^i_{jk}$, which corresponds to the component $\nabla^{g^\perp}$. When this tensor vanish identically the holonomy is $G$ or a smaller subgroup, as the connection $\nabla^g$ will coincide with the Levi-Civita connection. Heuristically, the torsion measures the failure for the holonomy for being $G$.

The torsion $T_{ijk}$ will play a significant role in the following discussion and it may be instructive to give an explicit example. Let us recall that in four dimensions the isomorphism $SO(4)\simeq SU(2)_L\times SU(2)_r$ induces the decomposition $6\to 3 + 3$ of a Maxwell tensor $F_{ab}$ into self-dual and anti-self dual components. Consider now the analogous  for the group $G=G_2$ discussed above. An antisymmetric tensor $A_{ab}$ transform as the adjoint of group $SO(7)$, which has $21$ generators, and the embedding of $G_2$ into $SO(7)$ induce the decomposition $21\to 14 + 7$ of $A_{ab}$, with $14$ corresponding to the adjoint and $7$ to the fundamental representation of $G_2$. This implies that $A_{ab}$ can be decomposed as
\be\lb{147}
A_{ab}=A_{ab}^+ + A_{ab}^-,
\ee
corresponding to $14$ and $7$ respectively. These components are explicitly
\be\lb{deverv}
A_{ab}^+=\frac{2}{3}(A_{ab}+\frac{1}{4}c_{abcd}A_{cd}),
\ee
\be\lb{deverv2}
A_{ab}^-=\frac{1}{3}(A_{ab}-\frac{1}{2}c_{abcd}A_{cd}).
\ee
In particular, the spin connection $\omega_{ab}$ of a given 7-dimensional metric can be expressed as $\omega_{ab}=(\omega_{ab})_++(\omega_{ab})_-$ in the same way as (\ref{147}). This induce a decomposition of the form (\ref{schemo}) for the Levi-Civita connection, the torsion part being related to  $(\omega_{ab})_-$. When this component is zero, then the torsion will also vanish and the holonomy will be in $G_2$.

Although the torsion may be interpreted as an obstruction of the holonomy to be reduced, the following detail should be remarked. Even in the case when the forms $\sigma_p^G$ corresponding to a $G$ structure are not covariantly constant, it may be incorrect to conclude that the holonomy is not reduced. As there is a local $SO(n)$ freedom for choosing the frame $e^a$, it may be the case that by a suitable rotation of the $e^a$ one may construct a new $G$ structure corresponding to the same metric and which, in addition, is covariantly constant. Thus the holonomy will be $G$ although the initial structure was not preserved by the Levi-Civita connection. An useful criteria for deciding whether or not a given metric is of $G$ holonomy is the fact that metrics with reduced holonomy are always Ricci flat or Einstein. This criteria is independent on the choice of the $G$ structure.

In addition to the $G_2$ case discussed above, other well known example of Ricci flat manifolds of reduced holonomy are hyperkahler ones, which encode several non compact gravitational instantons and also $K_3$ surfaces. By definition a hyperkahler manifold is $4n$ dimensional and admits a metric $g_{4n}$ whose holonomy group is in $Sp(n)$. For these manifolds there always exist a triplet $J_i$ (i=1,2,3) of $(1,1)$ tensors with quaternion multiplication rule $J^i J^j=\delta_{ij}I+\epsilon_{ijk}J^k$ such that the metric is hermitian with respect to any of them. The Lie algebra $sp(n)$ of $Sp(n)$ is generated by $(1,1)$ tensors $A$ of $so(4n)$ which commute with the $J^i$, i.e, satisfying $[A, J_i]=0$. In other words the action of $Sp(n)$ leave the tensors $J_i$ invariant. The generalization of the discussion given in the previous paragraph implies that when
\be\lb{efvr}
 \nabla_X J^i=0,
 \ee
 the holonomy will be included in $Sp(n)$. The last formula together with $\nabla_X g=0$ imply that the $Sp(n)$ invariant 2-forms $\omega_i(X,Y)=g(X, J^i Y)$ are also covariantly constant with respect to the Levi-Civita connection. These are known as Kahler forms, and this condition implies that the metric is Kahler with respect to any of the $\omega_i$. It can be shown that this system is equivalent to $d\omega_i=0$, and the $\omega_i$ together with the metric $g_{4n}$ compose the $Sp(n)$ structure.

For the Einstein case the classical examples are quaternion Kahler manifolds of dimension higher than four, which are $4n>4$ dimensional manifolds endowed with a metric $g_{4n}$ whose holonomy is in $Sp(n)\times Sp(1)\in SO(4n)$ \cite{Ishiharaq}-\cite{Ishiharaq2}. The set $J_i$ together with the set $A$ satisfying that $[A, J_i]=0$  are the generators of the Lie algebra $sp(n)\oplus sp(1)$, and the action of $Sp(n)$ leave the $J^i$ invariant but the action of $Sp(1)$ mix them due to the non trivial commutator $[J^i, J^j]=\epsilon_{ijk}J^k$. As a result if the condition
\be\lb{qker}
\nabla_X J^i=\epsilon_{ijk}J^k\widetilde{\omega}_-^k,\qquad \nabla_X \omega_i=\epsilon_{ijk}\omega_j\;\widetilde{\omega}_-^k,
\ee
then the manifold will have holonomy in $Sp(n)\times Sp(1)$. Here $\widetilde{\omega}_-^k$ is the $Sp(1)$ part of the connection. In different way than for hyperkahler manifolds, in the quaternionic case the triplet of 2-forms $\omega_i$ are not covariantly constant.  Still their specific behaviour (\ref{qker}) imply a reduction of the holonomy from $SO(4n)$ to $Sp(n)\times Sp(1)$. Alternatively it may be shown that the condition for being quaternion Kahler imply that
\be\lb{qker2}
d\omega^i=\epsilon_{ijk}\omega^j\wedge\widetilde{\omega}_-^k, \qquad d\Omega=0,
\ee
where the four form $\Omega=\omega_1 \wedge \omega_1+\omega_2\wedge \omega_2+\omega_3\wedge \omega_3$ has been introduced.  The $Sp(n)\times Sp(1)$ structure is composed by the metric, the three 2-forms $\omega_i$ and the four form $\Omega$.

\subsection{A check of the Papadopoulos list}

The present subsection deals with the problem of classifying which $G$ structures do admit some of their $G$ invariant p-forms $\sigma_p^G$ as a Killing-Yano tensors. This was investigated already in \cite{Papadopoulos1} with $G$ being the Berger groups. The purpose of the present section is to reproduce by use of the torsion languages developed in \cite{Agrikola1}-\cite{Karigiannis}. The Killing-Yano condition (\ref{cherri}) is translated for $\sigma_p^G$ as
\be\lb{kver}
\nabla_X\sigma^g_p=\frac{1}{p+1}i_X d\sigma^g_p.
\ee
All the forms $\sigma_p^g$ composing a Ricci flat structure are Killing-Yano, as both the left and the right hand side vanish identically. Our task is to find non trivial examples. The left hand side of the last equation involves the torsion $T^i_{jk}$ introduced (\ref{schemo}). The right hand is also determined in terms of $T_{ij}^k$ by the well known formula
\be\lb{covdif}
d\Lambda=\frac{1}{(p-1)!}\nabla_{[\mu_1}\Lambda_{\mu_2...\mu_p]}dx^{1}\wedge...\wedge dx^p,
\ee
together with (\ref{schemo}). Therefore the Killing-Yano equation is essentially reduced to a constraint for the torsion. The interesting point is that several solutions for these constraints involve structures which are relevant for constructing supergravity solutions with conformal field theory duals. The task to find hidden symmetries in these structures is therefore of theoretical interest.

\subsubsection{Kahler and Calabi-Yau structures}

Consider first $U(n)$ structures, which are defined in $d=2n$ dimensions \cite{GrayHervella}. These are composed by a 2n-dimensional metric $g$ defined over a manifold $M_{2n}$ and an almost complex structure $J$. The last is an automorphism of the cotangent space satisfying the complex imaginary unit multiplication rule $J^2=-I_{2n\times 2n}$, and the metric $g$ is assumed to be hermitian with respect to it. The hermiticity condition means that the tensor $\omega(X, Y)=g(X, J, Y)$ is a two form, commonly known as almost kahler form. The Nijenhuis tensor corresponding to $J$ may be expressed as
\be\lb{nijui}
N_{\mu\nu}^{\rho}=J_{\mu}^{\lambda}(\partial_{\lambda}J_{\nu}^{\rho}-\partial_{\nu}J_{\lambda}^{\rho})
-J_{\nu}^{\lambda}(\partial_{\lambda}J_{\mu}^{\rho}-\partial_{\mu}J_{\lambda}^{\rho}),
\ee
and the vanishing of this tensor implies that $M_{2n}$ is complex with respect to $J$. If in addition there exists a connection $\nabla^{u(n)}$ with torsion for which $\nabla^{u(n)} g=\nabla^{u(n)} J=0$ then the Nijenhuis tensor may be expressed entirely in terms of $J$ and the torsion. This condition is explicitly
\be\lb{covdev}
\nabla^{u(n)}_{\mu} J_{\nu}^{\rho}=\partial_{\mu}J_{\nu}^{\rho}+\gamma_{\lambda\mu}^{\rho} J_{\nu}^{\lambda}-\gamma_{\nu\mu}^{\lambda} J_{\lambda}^{\rho}=0,
\ee
with $\gamma_{\mu\nu}^{\rho}$ defined in terms of the Christoffel symbols $\Gamma_{\mu\nu}^{\rho}$ and the torsion $T_{\mu\nu}^{\rho}$ as follows
\be\lb{karv}
\gamma_{\mu\nu}^{\rho}=\Gamma_{\mu\nu}^{\rho}-\frac{1}{2}T_{\mu\nu}^{\rho}.
\ee
From (\ref{covdev})-(\ref{karv}) it follows that (\ref{nijui}) can be expressed as
\be\lb{puper}
N_{\mu\nu}^{\rho}=T_{\mu\nu}^{\rho}-J_{\mu}^{\lambda}J_{\nu}^{\sigma}T_{\lambda\sigma}^{\rho}+
(J_{\mu}^{\lambda}T_{\lambda\nu}^{\sigma}-J_{\nu}^{\lambda}T_{\lambda\mu}^{\sigma})J_{\rho}^{\sigma},
\ee
which express the Nijenhuis tensor entirely in terms of the torsion and the almost complex structure.
The decomposition $so(2n)=u(n)\oplus u(n)^{\bot}$ induce a decomposition of the space $\Lambda^2$ of two forms on $M_{2n}$ as
\be\lb{dikonpo}
\frac{1}{2}m(m-1)\longrightarrow m^2\oplus \frac{1}{2}m(m-1)\oplus\overline{\frac{1}{2}m(m-1)}.
\ee
This can be expressed as $\Lambda^2=\Lambda^{(1,1)}\oplus \Lambda^{(2,0)+(0,2)}$. Denote as $\Upsilon_{ijk}$ the following covariant derivatives
\be\lb{explota}
\nabla_i \omega_{jk}= \Upsilon_{i,jk}.
\ee
The torsion belongs to $T^{\ast}M \otimes u(n)^{\bot}$ and by representing the cotangent space as $T^{\ast}M=T^{\ast(1,0)}M\oplus T^{\ast(0,1)}M$ and taking into account (\ref{dikonpo}) it follows that the non zero covariant derivatives are
\be\lb{tuzhur}
\Upsilon_{\alpha,\beta\gamma},\qquad \Upsilon_{\alpha,\overline{\beta}\overline{\gamma}},\qquad\Upsilon_{\overline{\alpha},\beta\gamma},\qquad \Upsilon_{\overline{\alpha},\overline{\beta}\overline{\gamma}}.
\ee
These components can be divided into four irreducible representations $W_i$ with $i=1,..,4$ of $T^{\ast}M \otimes u(n)^{\bot}$ on $u(n)$ given by \footnote{Useful formulas for physicist related to these structures can be found in \cite{Papadopoulos2} and references therein.}
$$
(W_1)_{\overline{\alpha}\overline{\beta}\overline{\gamma}}=\Upsilon_{[\overline{\alpha},\overline{\beta}\overline{\gamma}]},\qquad (W_2)_{\overline{\alpha}\overline{\beta}\overline{\gamma}}=\Upsilon_{\overline{\alpha},\overline{\beta}\overline{\gamma}}
-\Upsilon_{[\overline{\alpha},\overline{\beta}\overline{\gamma}]}
$$
\be\lb{maders}
(W_3)_{\overline{\alpha}\beta\gamma}=\Upsilon_{\overline{\alpha},\beta\gamma}
-\frac{2}{m-1}\Upsilon_{\overline{\mu}},^{\overline{\mu}}_{[\gamma} g_{\beta]\overline{\alpha}},
\qquad (W_4)_{\gamma}=\Upsilon_{\overline{\mu}},^{\overline{\mu}}_{\gamma}.
\ee
The last component $W_4$ is traceless.

For $SU(n)$ structures one has, in addition to the Kahler form $\omega$, an invariant $(n,0)$ form $\Omega$ whose square is proportional to the volume form of $g$ namely
\be\lb{fveri}
(-1)^{\frac{m(m-1)}{2}}\bigg(\frac{i}{2}\bigg)^n\;\Omega\wedge\overline{\Omega}=\frac{1}{n!}\;\omega^n= dvol(g).
\ee
The additional covariant derivative
\be\lb{novlca}
\nabla_{\overline{\alpha}}\Omega_{\beta\gamma\delta\rho}=(W_5)_{\overline{\alpha}\beta\gamma\delta\rho},
\ee
determines a new class $W_5$.

The case $SU(3)$ is of particular importance in the context of compactifications of II supergravity down to four dimensions. These structures are classified as follows. As the components of the Nijenhuis tensor are expressed in terms $W_1$ and $W_2$ if these torsion components are zero the manifold is complex. Particular subcases are structures for which the unique non vanishing classes are $W_3$ and $W_5$ which are known as balanced. When $W_3$ is the unique non vanishing torsion the structure is known as special hermitian, when $W_5$ is the only non vanishing component the structure is Kahler. Another important examples are those for which $\partial \overline{\partial} J=0$ and $dJ\neq 0$ which are known as strong Kahler structures. If instead $W_1$ or $W_2$ are not zero, then the manifold is non complex. When $W_1$ is the unique non zero component the manifold is known as nearly kahler. When $W_2$ is the unique non zero component then the manifold is known as almost Kahler. Finally when the torsion belongs to $W_1^-\oplus W_2^-\oplus W_3$ the manifold is known as half flat.

The torsion classes $W_i$ not only determine the covariant derivatives of  $\omega(X,Y)$ and of $\Omega$, but also their differentials $d\omega$ and $d\Omega$ \cite{Chiossi}. This follows from the elementary formula (\ref{covdif}) and the final result for $SU(n)$ structures may be schematically stated as
$$
W_1\qquad \longleftrightarrow\qquad  d\omega^{(3,0)}+d\omega^{(0,3)},
$$
$$
W_3+W_4\qquad \longleftrightarrow\qquad d\omega^{(2,1)}+d\omega^{(1,2)},
$$
\be\lb{ber}
W_1+W_2\qquad \longleftrightarrow\qquad d\Omega^{(n-1,2)}+d\Omega^{(2,n-1)},
\ee
$$
W_4+W_5\qquad\longleftrightarrow\qquad d\Omega^{(n,1)}+d\Omega^{(1,n)}
$$
For example, the first (\ref{ber}) is explicitly
\be\lb{chcu}
d\omega^{(3,0)}+d\omega^{(0,3)}=3 W_1.
\ee
By comparing this formula with the first (\ref{maders}) it follows easily that when $W_2=W_3=W_4=W_5=0$ one has that
\be\lb{ferenz}
\nabla_X\omega=\frac{1}{3}i_X d\omega,
\ee
and this implies that for these types of manifolds the almost Kahler form $\omega$ is a Killing-Yano tensor. As it was discussed above, structures
with this types of torsion are nearly Kahler \cite{Graynearlykahler}. These manifolds are characterized by the condition $\nabla_X J (X)=0$ and some applications in physics can be found in \cite{Friedrich2}-\cite{Lust}. Additionally the last (\ref{ber}) together with the definition (\ref{novlca}) shows that when $W_4=W_5=0$ the components $\Omega^{(n,1)}$ and $\Omega^{(1,n)}$ are covariantly constant, thus Killing-Yano. These structures are balanced and hermitian.

\subsubsection{Quaternion Kahler and hyperkahler structures}

The next structures to consider are $Sp(n)\times Sp(1)$ ones, which are known as quaternion kahler. In this case the p-forms defining the structure are the triplet of almost kahler 2-forms $\omega_i$ together with the 4-form
\be\lb{4}
\Omega=\omega_1\wedge \omega_1+\omega_2\wedge \omega_2+\omega_3\wedge \omega_3.
\ee
For this type of structure there exist an useful formula derived in the proposition 4.3 of the reference \cite{cabrera} which relate the covariant derivatives of the almost kahler forms $\omega_i$ with their differentials. The explicit form of this formula is
$$
\nabla_X \omega_1(Y, Z)=d\omega_1(X, Y, Z)-d\omega_1(X, J^1 Y, J^1 Z)+ d\omega_2(J^2X,Y, Z)+ d\omega_2(J^2X,J^1Y, Z)
$$
\be\lb{tklass}
+ d\omega_2(J^2X,Y, J^1Z)+d\omega_3(J^2 X, Y, Z)-d\omega_3(J^2X, J^1Y,  J^1Z),
\ee
and the analogous formula holds for cyclic permuted indices. Clearly, if $\omega_1$ is required to be Killing-Yano then (\ref{kver})
implies that $d\omega_2=d\omega_3=0$. Furthermore $d\omega_1(X, J^1 Y, J^1 Z)=0$ and thus $d\omega_1(X,  Y, Z)$ will also vanish, as the map $J_i$ does not have kernel. The same reasoning follows for $\omega_2$ and $\omega_3$, thus these forms are also closed. Therefore if these forms are Killing-Yano then the metric is hyperkahler and the holonomy is $Sp(n)$ or a subgroup. In addition the covariant derivatives of the four form (\ref{4}) can be calculated by direct use of (\ref{tklass}), the result is given in the formula (5.1) of the reference \cite{cabrera2}
\be\lb{qk}
\nabla\Omega=2\epsilon_{ijk}\alpha_{[kj]}\wedge \omega_i.
\ee
In (\ref{qk}) the tensor $\alpha_{kj}$ is defined as
\be\lb{ves}
\alpha_{kj}(X,Y,Z)=\alpha_{k}(X, J^j Y, Z),
\ee
with
$$
\alpha_i=-\lambda_i\otimes g+\frac{\epsilon_{ijk}}{4}\bigg(\nabla \omega_k(\cdot, J^j\cdot, \cdot)-\nabla \omega_k(\cdot, \cdot, J^j\cdot)\bigg).
$$
In the last formula the 1-forms $\lambda_i$ are defined as
$$
\lambda_1(X)=\frac{1}{2n}<\nabla_X\omega_2, \omega_3>,
$$
up to cyclic permutations. We see from (\ref{qk}) $\Omega$ is a Killing-Yano tensor when $\Omega$ is covariantly constant in general. Thus the metric is quaternion Kahler in a generic case.

\subsubsection{$Spin(7)$ structures}

 Other examples with physical interest are the $Spin(7)$ structures \cite{FernandezSpin}. In this case the manifold is eight-dimensional with metric $g_8=\delta_{ab}e^a \otimes e^b$. The holonomy is in $Spin(7)$ if the following octonionic form
\be\lb{octur}
\Phi=e^8\wedge \phi+\ast_7 \phi,
\ee
is closed. Here $\phi=c_{abc}e^a\wedge e^b\wedge e^c$ and $c_{abc}$ are the multiplication constants of the imaginary octonions. This form satisfy the self duality condition $\ast \Phi=\Phi$. In addition we have that
\be\lb{Spin7}
d\Phi=\theta\wedge \Phi+W_1,
\ee
that is the differential of $\Phi$ has a part which is proportional to $\Phi$ and a part $W_1$ which is not.
The form $\theta$ is known as the Lee form. The covariant derivative of the fundamental four form  is
\be\lb{spino}
\nabla_m\Phi_{ijkl}=T_{mip}g^{pq}\Phi_{qjkl}+T_{mjp}g^{pq}\Phi_{iqkl}+T_{mkp}g^{pq}\Phi_{jiql}+T_{mlp}g^{pq}\Phi_{jklq},
\ee
with $T$ given by
\be\lb{ilya}
T=-\ast d\Phi-\frac{7}{6}\ast(\theta\wedge \Phi)
\ee
By checking explicitly the condition (\ref{kver}) in this situation we were able to find a solution only when $d\Phi=\nabla\Phi=0$ and the manifold has holonomy in $Spin(7)$.

\subsubsection{$G_2$ structures}

Let us consider what happens for a $G_2$ structure ($\phi$, $\ast\phi$) \cite{Grayweakholonom}, \cite{FernandezG2}. In this case, one may find non trivial Killing-Yano tensors, as it will be seen below. The torsion classes $\tau_i$ for the differential are given by \cite{Chiossi}
\be\lb{G2}
d\phi=\tau_0\ast \phi+3\tau_1\wedge \phi+\ast\tau_2,
\ee
$$
d\ast\phi=4\tau_1\wedge \ast\phi+\ast\tau_3.
$$
When the torsion classes vanish the holonomy will be $G_2$ or a subgroup of $G_2$. The covariant derivative of the three form can be expressed as \cite{Karigiannis}
\be\lb{cvG2}
\nabla_l\phi_{abc}=T_{lm}g^{mn}(\ast\phi)_{nabc}
\ee
with the torsion tensor given by
\be\lb{torter}
T_{lm}=\frac{\tau_0}{4}g_{lm}-(\tau_3)_{lm}+(\tau_1)_{lm}-(\tau_2)_{lm}.
\ee
The Killing-Yano condition $4\nabla_X\phi=i_X d\phi$ implies $i_X \nabla_X \phi=0$. This together with (\ref{cvG2})-(\ref{torter}) shows that for $\phi$ being a Killing-Yano tensor only a non zero $\tau_0$ component is allowed. These structures are known as nearly parallel. Thus for every nearly parallel $G_2$ structure the octonionic three form $\phi$ is a non trivial Killing-Yano tensor of order three.

\subsection{Further examples}
\subsubsection{Almost contact structures}
The calculations performed above show the validity of the Papadopoulos list for Killing-Yano tensors in $G$ structures of the Berger type. Below we will be focused on cases which are not of this type, and which do not appear in the Papadopoulos list. Such is the case for the almost contact structures.

Almost contact structures are defined in $d=2n+1$ dimensions and are intimately ligated to almost Kahler structures in dimension $d=2n+2$. In fact the cone of an almost contact structure defines an almost Kahler structure and when the structure is Kahler the almost contact structure is known as sasakian. Sasakian structures are reviewed for instance in \cite{boyer1}-\cite{boyer6}. When the Kahler cone metric is Ricci flat, thus Calabi-Yau, then the odd dimensional metric is known as Einstein-Sasaki.

In formal terms, an almost contact structure is a $U(n)\times 1\in SO(2n+1)$ structure. It is composed by a metric $g_{2n+1}$ defined over a space $M_{2n+1}$ together with a selected vector field $\xi \in TM_{2n+1}$ whose dual form will be denoted as $\eta\in T^{\ast}M_{2n+1}$, and a morphism $\phi:TM_{2n+1}\to TM_{2n+1}$ satisfying the conditions
$$
g_{2n+1}(\phi X, \phi Y)=g_{2n+1}(X,Y)-\eta(X)\eta(Y),
$$
\be\lb{blue2}
\phi^2=-I+\eta\otimes \xi.
\ee
The fundamental form for this structure is $\Phi=g_{2n+1}(X,\phi Y)$. The cone over an almost contact structure
\be\lb{conus}
g_{2n+2}=dr^2+r^2 g_{2n+1},
\ee
is defined over $M_{2n+2}=R_{>0}\times M_{2n+1}$ and admits an almost complex structure $J$ over $M_{2n+2}$ described by the following actions
\be\lb{almcom}
J \partial_r=-\frac{1}{r}\xi,\qquad J X=\phi X+r \eta(X) \partial_r.
\ee
By decomposing a vector field $\widetilde{X}\in R_{>0}\times M_{2n+1} $ into a radial and angular part as $\widetilde{X}=(a, X)$, it is found
from (\ref{blue2}) and (\ref{almcom}) that the action of the almost complex structure over $\widetilde{X}$ is
\be\lb{axion}
J(a, X)=(r\eta(X), \phi X-\frac{a}{r}\xi).
\ee
The lifted Levi-Civita connection $\widetilde{\nabla}$ over the cone is defined through
$$
\widetilde{\nabla}_{\partial_r}\partial_r=0,\qquad \widetilde{\nabla}_{X}\partial_r=\widetilde{\nabla}_{\partial_r}X=\frac{X}{r}
$$
\be\lb{impous}
\widetilde{\nabla}_X Y=\nabla_X Y-r g(X,Y)\partial_r.
\ee
Here $\nabla$ is the Levi-Civita connection for the metric $g_{2n+1}$ of the almost contact structure. From (\ref{impous}) and (\ref{almcom}) it is deduced that
$$
(\widetilde{\nabla}_{\partial_r} J)\partial_r=(0,0),\qquad (\widetilde{\nabla}_{\partial_r}) X=(0,0),
$$
\be\lb{simama}
(\nabla_X J)\partial_r=(0,\frac{1}{r}(-\nabla_X \xi+\phi X))
\ee
$$
(\widetilde{\nabla}_{X} J)Y=(r \nabla_X \eta(Y)-r g_{2n+1}(X,\phi Y), (\nabla_X\phi)Y-g_{2n+1}(X,Y)\xi+\eta(Y)X).
$$
The Kahler condition is equivalent to the vanishing of the all the covariant derivatives (\ref{simama}) and this holds when
$$
\nabla_X \xi=\phi X
$$
\be\lb{Ensak}
\nabla_X \eta(Y)=g_{2n+1}(X,\phi Y),
\ee
$$
(\nabla_X\phi)Y=g_{2n+1}(X,Y)\xi-\eta(Y)X.
$$
The last three conditions define a Sasakian structure. Alternatively the second (\ref{Ensak}) implies that $\xi$ is Killing and the first and the third ones may combined to obtain that
\be\lb{saskon}
\nabla_X(d\xi^\ast)=-2 X^{\ast}\wedge \xi^{\ast}.
\ee
This equation imply in particular that $d^{\ast}d\xi^{\ast}=(n-1)\xi^{\ast}$ and taking this into account it follows that (\ref{saskon})
can be expressed in the following manner
\be\lb{saskon2}
\nabla_X(d\xi^\ast)=-\frac{1}{n-1} X^{\ast}\wedge d^{\ast}d\xi^{\ast}.
\ee
As $d\xi^{\ast}$ the last equation shows that $d\xi^\ast$ is a conformal Killing tensor \cite{Sewemann2}. This, together with the fact that $\xi^{\ast}$ is also a conformal Killing 1-form, implies that the combinations
\be\lb{cob}
\omega_k=\eta\wedge (d\eta)^k,
\ee
are all Killing tensors of order $2k+1$ \cite{Sewemann2}.

There exist other almost contact structures, different from Sasaki ones, and which also admit Killing-Yano tensors. Generic almost contact structures are characterized by the irreducible components of the covariant derivative $\nabla \Phi$ of the fundamental form \cite{chinea}-\cite{chinea2}. In representation theoretical terms this derivative belongs to $T^{\ast}M\otimes u(n)^{\perp}$. One may decompose the cotangent space as
\be\lb{dik1}
T^{\ast}M=R \eta+\eta^{\perp},
\ee
from where it follows that
\be\lb{dik2}
so(2n+1)\simeq \Lambda^2 T^{\ast}M=\Lambda^2 \eta^{\perp}+\eta^{\perp}\wedge R \eta
\ee
$$
=u(n)+u(n)^{\perp}_{\mid \xi^{\perp}}+\eta^{\perp}\wedge R \eta.
$$
From (\ref{dik2}) it is obtained that
\be\lb{dik3}
u(n)^{\perp}=u(n)^{\perp}_{\mid \xi^{\perp}}+\eta^{\perp}\wedge R \eta.
\ee
Therefore the covariant derivative $\nabla \Phi$ belongs to
\be\lb{nablapertenece}
\nabla \Phi \in T^{\ast}M\otimes u(n)^{\perp}= \eta^{\perp}\otimes u(n)^{\perp}_{\mid \xi^{\perp}}+\eta\otimes u(n)^{\perp}_{\mid \xi^{\perp}}
\ee
$$
+\eta^{\perp}\otimes \eta^{\perp}\wedge \eta+\eta\otimes \eta^{\perp}\wedge \eta.
$$
The respective components are
\be\lb{respcomp}
\nabla_i \phi_{jk},\qquad \nabla_m \phi_{jk},\qquad \nabla_i \phi_{mk},\qquad \nabla_m \phi_{mk}.
\ee
with the indices $i, j, k$ corresponding to the $\eta^{\perp}$ directions and $m$ to the $\eta$ direction. But these components are not irreducible, and in fact it was shown in \cite{chinea}-\cite{chinea2} that there is a further decomposition into 12 irreducible classes given schematically as
$$
\eta^{\perp}\otimes u(n)^{\perp}_{\mid \xi^{\perp}}=C_1+C_2+C_3+C_4
$$
$$
\eta^{\perp}\otimes \eta^{\perp}\wedge \eta=C_{5}+C_{6}+C_{7}+C_{8}+C_{9}+C_{10}
$$
\be\lb{chinogon}
\eta\otimes u(n)^{\perp}_{\mid \xi^{\perp}}=C_{11},\qquad
\eta\otimes \eta^{\perp}\wedge \eta=C_{12}.
\ee
More precisely, the space $C(V)$ of 3-three tensors with the same symmetries of $\nabla \Phi$ is
$$
C(V)=\{T\in \otimes_3 V\mid T(x, y, z)=-T(x, z, y)=-T(x, \phi y, \phi z)+\eta(y)T(x,\xi, z)+\eta(z) T(x, y, \xi)\},
$$
and can be decomposed as
$$
C(V)=\bigoplus_{i=1}^{12}C_i(V)
$$
with the irreducible components $C_i(V)$ given by
$$
C_1(V)=\{T\in C(V)\mid T(x, x, y)=-T(x, y, \xi)=0\}
$$
$$
C_2(V)=\{T\in C(V)\mid T(x, y, z)+T(y, z, x)+T(z, x, y)=0, \;\; T(x,y,\xi)=0\}
$$
$$
C_3(V)=\{T\in \otimes_3 V\mid T(x, y, z)=T(\phi x, \phi y, z), \;\; \sum c_{12}T(x)=0\}
$$
$$
C_4(V)=\{T\in \otimes_3 V\mid T(x, y, z)=\frac{1}{2n-1}\bigg[(g(x, y)-\eta(x)\eta(y))c_{12}T(z)
$$
$$
-\frac{1}{2n-1}(g(x, z)-\eta(x)\eta(z))c_{12}T(y)-g(x, \phi y)c_{12}T(\phi z)
$$
$$
+g(x, \phi z)c_{12}T(\phi y)\bigg],\;\; c_{12}T(\xi)=0\}
$$
$$
C_5(V)=\{T\in \otimes_3 V\mid T(x, y, z)=\frac{1}{2n}\bigg[g(x, \phi z)\eta(y)\overline{c}_{12}T(\phi \xi)-g(x, \phi y)\eta(z)\overline{c}_{12}T(\phi \xi)\bigg]\}
$$
$$
C_6(V)=\{T\in \otimes_3 V\mid T(x, y, z)=\frac{1}{2n}\bigg[g(x, y)\eta(z)c_{12}T(\xi)-g(x, z)\eta(y)c_{12}T(\phi \xi)\bigg]\}
$$
$$
C_7(V)=\{T\in \otimes_3 V\mid T(x, y, z)=T(y, x,\xi)\eta(z)-T(\phi x, \phi z, \xi)\eta(y),\;\;c_{12}T(\xi)=0\}
$$
$$
C_8(V)=\{T\in \otimes_3 V\mid T(x, y, z)=-T(y, x,\xi)\eta(z)-T(\phi x, \phi z, \xi)\eta(y),\;\;\overline{c}_{12}T(\xi)=0\}
$$
$$
C_9(V)=\{T\in \otimes_3 V\mid T(x, y, z)=T(y, x,\xi)\eta(z)+T(\phi x, \phi z, \xi)\eta(y)\}
$$
$$
C_{10}(V)=\{T\in \otimes_3 V\mid T(x, y, z)=-T(y, x,\xi)\eta(z)+T(\phi x, \phi z, \xi)\eta(y)\}
$$
$$
C_{11}(V)=\{T\in \otimes_3 V\mid T(x, y, z)=-T(\xi, \phi y, \phi z)\eta(x)\}
$$
$$
C_{12}(V)=\{T\in \otimes_3 V\mid T(x, y, z)=-T(\xi, \xi, z)\eta(x)\eta(y)-T(\xi, y, \xi)\eta(x)\eta(z)\}
$$
where the following quantities
$$
c_{12}T(x)=\sum T(e_i, e_i, x)
$$
$$
\overline{c}_{12}T(x)=\sum T(e_i, \phi e_i, x)
$$
has been introduced, with $e^i$ an arbitrary orthonormal basis.

It shoule be remarked that some of the classes $C_i$ may vanish for lower enough dimensions. For $n=1$ the covariant derivative $\nabla \Phi$ belongs to $C_5\oplus C_6\oplus C_9\oplus C_{12}$. The case $n=2$ corresponds to the structures studied in \cite{salaconti}-\cite{salaconti2}, and for this dimensions almost contact structures belongs to $C_2\oplus C_4\oplus C_6\oplus C_8\oplus C_{10}\oplus C_{12}$. Only for $n\geq 3$ all the classes may not vanish.

The classification of the structures goes as follows. When all the classes vanish the structure is known as cosympletic, $C_1$ structures are nearly-K-cosympletic, $C_5$ are $\alpha$-Kenmotsu manifolds, $C_6$ are $\alpha$-Sasakian and in particular, Sasakian structures belong to this class. Other structures are $C_5\oplus C_6$ which are known as trans-sasakian, $C_2\oplus C_9$ which are almost cosympletic, $C_6\oplus C_7$ which are quasi-sasakian, $C_1\oplus C_5\oplus C_6$ which are nearly trans-Sasakian and $C_1\oplus C_2\oplus C_9\oplus C_{10}$ which are quasi K-cosympletic and $C_3\oplus C_4\oplus C_5\oplus C_6 \oplus C_7 \oplus C_8$ which are normal ones. Properties of these structures may be found in \cite{cabre}-\cite{congo} and references therein.

The class for which $\Phi$ is Killing-Yano is the one for which $\nabla \Phi$ is totally anti-symmetric, and this is the case when the unique non vanishing class is $C_1$. Therefore for nearly K-cosympletic structures are the ones for which the fundamental form $\Phi$ is a Killing-Yano tensor of order two. These structures are characterized by the condition $\nabla_X \phi X=0$, i.e, $\nabla_X \phi Y+\nabla_Y \phi X=0$. Properties of this structures were studied for instance in \cite{nerlco1}-\cite{nerlco12}.

\subsubsection{SO(3) structures in SO(5) and higher dimensional generalizations}

Let us consider now $SO(3)$ structures in five dimensions. Given a 5-dimensional manifold $M_5$ with a metric $g_5$ an $SO(3)$ structure is the reduction of the frame bundle to a $SO(3)$ sitting in $SO(5)$ \cite{Friedrso(3)}. One has the decomposition $so(5)=so(3)\oplus V$ with $V$ the unique 7-dimensional fundamental representation of $so(3)$. The space $R^5$ is isomorphic to space of $3\times 3$ symmetric traceless matrices $S^2_0R^3$, the isomorphism can be expressed by means of the mapping
\be\lb{eqovv}
(x_1, x_2, x_3, x_4, x_5)\qquad \longleftrightarrow\qquad X=\bigg(\begin{array}{ccc}
                               \frac{x_1}{\sqrt{3}}-x_4 & x_2 & x_3 \\
                               x_2 & \frac{x_1}{\sqrt{3}}+x_4 & x_5 \\
                               x_3 & x_5 & -\frac{2x_1}{\sqrt{3}}
                             \end{array}\bigg).
\ee
These matrices define the unique irreducible representation $\rho$ of $SO(3)$ in $R^5$ given as follows
\be\lb{irar}
\rho(h)X=h X h^{-1}, \qquad h\in SO(3).
\ee
For an element $X$ its characteristic polynomial $P_X(\lambda)$ invariant under the action of $\rho$, i.e, $P_{\rho(h)X}(\lambda)=P_X(\lambda)$, is given by
\be\lb{ivne}
P_X(\lambda)=\det(X-\lambda I)=-\lambda^3+g(X,X)\lambda+\frac{2\sqrt{3}}{9}\Upsilon(X,X,X),
\ee
with
$$
g(X, X)=x_1^2+x_2^2+x_3^2+x_4^2+x_5^2,
$$
$$
\Upsilon(X, X, X)=\frac{3\sqrt{3}}{2}\det X=\frac{x_1}{2}(6x_1^2+6x_2^2-2x_3^2-3x_4^2-3x_5^2)+\frac{3\sqrt{3}x_4}{2}(x_5^2-x_3^2)+3\sqrt{3}x_2 x_3x_5.
$$
By introducing a 3-tensor $\Upsilon_{ijk}$ by the relation $\Upsilon(X, X, X)=\Upsilon_{ijk}x_i x_j x_k$ it follows that
$$
\Upsilon_{ijk}=\Upsilon_{(ijk)},
$$
\be\lb{mog}
\Upsilon_{ijj}=0,
\ee
$$
\Upsilon_{jki}\Upsilon_{lni}+\Upsilon_{lji}\Upsilon_{kni}+\Upsilon_{kli}\Upsilon_{jni}=
g_{jk}g_{ln}+g_{lj}g_{kn}+g_{kl}g_{jn},
$$
where the tensor $g_{ij}$ is defined through the relation $g(X,X)=g_{ij} x_i x_j$. In these terms for a given manifold $M_5$ with a metric $g_5$ an $SO(3)$ structure is given in terms of a tensor $\Upsilon$ of rank three for which the associated linear map constructed in terms of $Z\in TM_5$ given by
$$
\Upsilon_{ij}=(\Upsilon_{ijk}Z_k)\in End(TM_5)
$$
satisfying the following conditions
$$
Tr(\Upsilon_Z)=0,
$$
\be\lb{endomorfo}
g(X, \Upsilon_Z Y)=g(Z, \Upsilon_Y X)=g(Y, \Upsilon_X Z),
\ee
$$
\Upsilon^2_Z Z=g(Z,Z)Z.
$$
There always exist a basis $e^a$ such that
\be\lb{coffe}
g_5(X, X)=e^1\otimes e^1+e^2\otimes e^2+e^3\otimes e^3+e^4\otimes e^4+e^5\otimes e^5,
\ee
which is defined up to an $SO(3)$ transformation $\widetilde{e}^a=\rho(h)e^a$, and such that
$$
\Upsilon=\frac{e^1}{2}\otimes(6e^1\otimes e^1+6e^2\otimes e^2-2e^3\otimes e^3-3e^4\otimes e^4-3e^5\otimes e^5)
$$
$$
+\frac{3\sqrt{3}e^4}{2}\otimes(e^5\otimes e^5-e^3\otimes e^3)
+3\sqrt{3}e^2\otimes e^3\otimes e^5.
$$
This define an $SO(3)$ structure for $(M_5, g_5)$.

The types of possible structures are defined in terms of the covariant derivative of $\Upsilon_{ijk}$. Different from the other $G$ structures considered above, this tensor is totally symmetric, and one may try to study in which situations $\Upsilon_{ijk}$ is a Killing tensor instead a Killing-Yano one. The Killing condition $\nabla_{(i}\Upsilon_{jkl)}$ can be rewritten as
\be\lb{kilos}
\nabla_X \Upsilon(X,X,X)=0.
\ee
Fortunately, structures satisfying this condition have been considered in \cite{Bobienski1}-\cite{Bobienski2} and we can just borrow the description from that references. The condition (\ref{kilos}) resembles the nearly Kahler one $\nabla_X J(X)=0$, and for this reason $SO(3)$ structures satisfying this condition are known as nearly integrable in the terminology of the references \cite{Bobienski1}-\cite{Bobienski2}. For example there exist only three nearly integrable structures with eight dimensional symmetry groups
$$
M_+=SU(3)/SO(3),\qquad M_0=(SO(3)\times_{\rho}R^5)/SO(3),\qquad M_-=SL(3,R)/SO(3).
$$
Further examples with symmetry group lower dimensional groups were found in \cite{Bobienski1}-\cite{Bobienski2} and on 5-dimensional Lie groups in \cite{fino}.

In addition to these examples, it was shown in \cite{Nurowski3} that tensors satisfying the conditions (\ref{mog}) exist in distinguished dimensions $n_k=3k+2$, where $k=1, 2, 4, 8$, as observed by Bryant. The numbers $k=1, 2, 4, 8$ are the dimensions of the division algebras and in these dimensions the orthogonal group may be reduced to the subgroups $H_k\subset SO(n_k)$, with $H_1=SO(3)$, $H_2=SU(3)$, $H_4=Sp(3)$ and $H_8=F_4$. Nearly integrable geometries can be defined in all these dimensions by the condition (\ref{kilos}) and it turns out that for all these geometries $\Upsilon$ is a Killing tensor. Examples of these geometries can be found in \cite{Nurowski3}.

\section{Discussion}

In the present work some of the applications of Killing-Yano tensors in General Relativity and Supersymmetric Quantum Field Theory has been reviewed.
Additionally the Papadopoulos list of $G$ structures whose $G$ invariant tensors are Killing-Yano has been reproduced and enlarged to cases which do not appear in the Berger list. It should be remarked that the results presented here about $G$ structures do not consist in a no go theorem. For instance, we have shown that between the $SU(3)$ structures, the nearly kahler are the ones for which their almost Kahler two form is Killing-Yano. But this does not imply the absence of Killing-Yano for other $SU(3)$ structures. In fact, the presence of Killing-Yano tensors in half-flat manifolds, which are outside this classification, are under current investigation \cite{doto}. What the present work is showing is that for these other structures the presence of a Killing-Yano tensor may be an special situation, while for the nearly Kahler case the presence of hidden symmetries is something generic. The same considerations hold for the other $G$ structures studied.

The presence of these hidden symmetries in these structures can be of interest in the AdS/CFT correspondence. For instance, we have shown that nearly Kahler, weak $G_2$ holonomy or Einstein-Sasaki manifolds do admit non trivial Killing-Yano tensors. The cones over these manifolds are Ricci flat and of holonomy $G_2$, $Spin(7)$ and $SU(3)$ holonomy respectively, and from these manifolds one may construct ten dimensional supergravity solutions whose near horizon limits are the form $AdS_3\times$(weak $G_2$), $AdS_4\times$(nearly Kahler) and $AdS_5\times$(Einstein-Sasaki). In some regimes certain anomalous dimensions of the dual quantum field theories may be calculated by studying strings configuration over these backgrounds. The energy and the conserved quantities for the movement of these strings give information about the anomalous dimensions of the dual theory. It is even possible to draw conclusions about the dual theory by studying particle limits of that string. For example in \cite{Kruczenski} anomalous long BPS operators are matched to massless point like strings in $AdS_5$ backgrounds with the Einstein-Sasaki spaces found in \cite{Martelli sparks} as internal spaces, and the conserved charges for that particle like movement gives information about the anomalous dimensions of that operators. Thus the presence of hidden symmetries in these backgrounds is of theoretical interest, and it may be an interesting task to understand to which quantum numbers of the dual theory these Killing-Yano tensors are matched with. Several of these are W-symmetries, as pointed out in \cite{Papadopoulos1}, but a more concrete description still is desirable.

Another interesting task could be to understand more deeply whether or not the relation between the algebraic type of the curvature for a given space time and the presence of hidden symmetries, which is known for Killing-Yano tensors of order two in four dimensions, can be generalized to higher dimensions and for tensors of higher rank. In our opinion, these task are of theoretical interest and deserve further attention.
\\

{\bf Acknowledgements:}  Discussions with I. Dotti, M. L. Barberis, G. Giribet, G. Silva, N. Grandi and M. Schvellinger are warmly acknowledged. When finished this paper, we were aware of the existence of the refernces \cite{LianaDavid}-\cite{Sewemann2} which may have overlap with the present work. The author is supported by CONICET (Argentina) and by the
ANPCyT grant PICT-2007-00849.
\\

\end{document}